\def \Liu {{\cal {L}}}
\def \vF  {{\bf {F}}}
\def \vA1  {\bf {A}_1}

\def \vB  {\bf {B}}
\def \vA  {\bf {A}}

\def \vri {{\bf {r}}_i}
\def \vr  {{\bf {r}}}
\def \vpi {{\bf {p}}_i}
\def \vp {{\bf {p}}}

\def \vv {{\bf {v}}}

\def \vecX {{\bf {X}}}
\def \vecY {{\bf {Y}}}
\def \vecn {{\bf {n}}}

\def \vecs {{\bf {S}}}
\def \vecd {{\bf {D}}}
\def \vcsi {{\bf {\xi}}}
\def \vw  {\bf {w}}
\def \grad {{\bf {\nabla}}}
\def \vecR  {{\bf {R}}}
\def \vecV  {{\bf {V}}}
\def \vb {{\bf {b}}}
\def \lb {{{\bar {\Omega}}}}
\def \vecn {{\bf {n}}}

\def \fop {{f^{\rm {op}}}}
\bigskip
\centerline {\bf DRAG ON  A SATELLITE MOVING ACROSS A SPHERICAL GALAXY}

\centerline {\bf I: TIDAL AND FRICTIONAL FORCES IN SHORTLIVED ENCOUNTERS}

\vskip 20pt 
\centerline {Monica  COLPI$^1$ and Andrea PALLAVICINI$^2$}

\bigskip

$^1$ Dipartimento di Fisica, Universit\`a Degli Studi di Milano, Milano, Italy

$^2$ Dipartimento di Fisica, Universit\`a di Pavia, Pavia, Italy

\vskip 30pt
\centerline {ABSTRACT}

The drag force on a satellite of mass $M$ moving with speed $\vecV$ 
in the gravitational field of a spherically symmetric background of 
stars 
is computed. 
During the encounter, the stars are subject
to a time-dependent force 
that alters their equilibrium.
The resulting distortion  in the stellar density field acts back to produce
a force $\vF_{\Delta}$  that decelerates the satellite.
This force is computed using a perturbative technique known as {\it linear 
response theory}.

In this paper, we extend the formalism of linear response
to  derive the correct expression for the back-reaction force 
$\vF_{\Delta}$ that applies when the stellar system 
is described by an equilibrium   {\it one-particle} distribution 
function. 
$\vF_{\Delta}$ 
is expressed in terms of  a 
suitable correlation function 
coupling the 
satellite dynamics 
to the unperturbed dynamics of the stars.
At time $t$ the force depends upon the whole history of the composite system.
In the formalism, we account for the shift 
of  the stellar center of mass resulting
from linear momentum conservation. The self-gravity of the 
response is neglected since it contributes  to a higher order 
in the perturbation.
Linear response theory applies also to the case of a satellite orbiting 
outside  the spherical galaxy.

The case of a satellite moving on a straight-line, 
at high speed relative to the 
stellar  dispersion velocity,   is explored.
We find that the satellite during its passage 
rises (a) {\it global tides} in the
stellar distribution and (b) a {\it wake}, i.e., an 
overdense region behind
its trail. 


If the satellite 
motion is external to 
the galaxy, 
it suffers a {\it dissipative} force which is not exclusively acting along 
$\vecV$ but acquires a component along $\vecR$, the position vector
relative to the 
center of the spherical galaxy.
We derive the analytical expression of the force, in the impulse 
approximation.

 In  penetrating  shortlived 
encounters, the satellite 
moves across the stellar distribution and 
the wake  excited in the density field
is responsible for most  of the deceleration.
We find that 
{\it dynamical friction}  rises from a {\it memory} effect involving only 
those stars perturbed along the path.
The force can be written in terms of an effective Coulomb logarithm
which now depends on the dynamical history, and in turn upon time $t$.
$\ln\Lambda$ is computed for two simple equilibrium density distributions:
It is shown that the drag increases as the 
the satellite approaches the denser regions of the stellar distribution. 
The  braking force then stays almost constant.
When 
the satellite crosses the edge of the galaxy the force 
does not vanish but declines  with time as $\vecR\to \infty.$ 

In the case of a homogeneous cloud, we compute the total energy loss.
In evaluating the contribution resulting from friction, we derive 
self-consistently  the maximum impact parameter which 
is  
found equal to the length travelled by the satellite within the system. 
Tides excited by the satellite  in the galaxy reduce the 
value of the energy loss by friction; in close  encounters
this value is decreased by a  factor $\sim  
1.5$. 

\noindent
{\it Subject headings:} galaxies: clustering -- stars:stellar dynamics 
 
\vskip 30pt

\centerline {I. INTRODUCTION}
\vskip 25pt

A massive object $M$ moving with velocity $\vecV$ through a background  of 
field stars  suffers a dissipative force known as dynamical friction.
In the original formulation of Chandrasekhar  (1943), 
this force is the  result of the momentum exchange between 
the particle  $M$ and the stars  
of the background.
In a uniform isotropic stellar system, the
uncorrelated superposition
of these binary encounters leads to a frictional  force 
$$\vF_{DF}=-4\pi {[GM]^2\over V^2}\rho(<V) \ln\left ({b_{max}
\over b_{min}}\right) {\vecV\over V}$$
where $\rho$ is the mass 
density of the stars having speed less than $V$, and 
$b_{max}$ and $b_{min}$ are the maximum and minimum impact parameters
for which encounters can be considered effective. 
$b_{max}$ is conventionally set equal to the size of the
stellar system surrounding the object (to avoid a divergence)
and
$b_{min}$ is the larger between the
characteristic radius of the satellite $M$ and the
impact parameter for a 90$^o$ scattering event (White 1976).
In the derivation of the above equation  the stars 
interact solely with the gravitational field
of the incoming particle and move
along trajectories that are straight lines at infinity.
Because of the isotropy of the surrounding stellar
field the force on $M$, responsible for its 
deceleration, is acting exclusively along
the direction of motion, i.e., along  $\vecV$.
Dynamical friction can  be viewed, alternatively,  as a 
force originating  from the overdensity excited in the stellar field by 
the particle in its motion: a wake develops 
behind $M$ which is causing its deceleration (Mulder 1983).

Dynamical friction plays a central role in many 
astrophysical environments:
It intervenes, for example, in the evolution of galaxies accreting small
dwarfs (Lin \& Tremaine 1983; Quinn \& Goodman 1986; Binney \& Tremaine 
1987), in galaxy cannibalism, in the collision 
and merger of galaxies 
(Barnes \& Hernquist 1992), and in the inspiral of a massive black hole at 
the center of dense stellar clusters (Begelman, Blandford \& Rees 1980). 
The expression of the frictional force in a homogenous background
has 
therefore been used to derive  simple 
analytical estimates
of the orbital decay times and to track evolutionary scenarios, for these cases.
In these  systems however, the ambient medium  is 
neither uniform, nor infinite and  this calls for a deep analysis of
this process.

Since the  work by Chandrasekhar, dynamical friction has been studied
using various techniques.
Tremai-
\break ne and Weinberg (1984), and Weinberg 
(1986) studied the  loss of angular momentum suffered by 
a satellite moving
in the gravitational field of a spherical galaxy,
 along  a circular orbit.  
Using a perturbative approach  
(that refers also to works by Lynden-Bell \& Kalnajs 1972)
for the description of  the interaction
of the satellite 
with the  background of self-gravitating point masses,
they found that angular momentum is exchanged 
secularly  with those stars whose orbits are commensurate, i.e.,  
in near-resonance with the satellite's orbit.
In applying the technique to the case of a satellite (modelled
as a Plummer  sphere) orbiting inside of 
a singular isothermal sphere, Weinberg (1986) found that the resulting torque
on $M$ is equivalent in magnitude to the one obtained using Chandrasekhar 
formula, applied to a locally homogeneous system: 
The maximum impact parameter 
 $b_{max}$ is found to be
comparable to the
size of the satellite orbit $R_s.$
This result  suggests   that the braking of a satellite 
depends more  
on the local properties of the stellar background, despite the
long-range nature of the gravitational force.

For a parent galaxy modelled as a polytrope and a satellite
as a non-deformable Plummer sphere 
similar findings  were  inferred by Bontekoe \& Van Albada (1987).
In a self-consistent $N$-body numerical calculation 
these authors observed that the decay process can still be
described as if it were a local phenomenon in which the density field
around the object determines primarily the magnitude of the
torque and in turn the decay time, for a satellite moving along 
a circular orbit (see also
Farouki, Duncan \& Shapiro 1983). 
Bontekoe \& Van Albada (1987) also explored the evolution of a satellite 
 moving around the primary galaxy on 
a grazing orbit, and  found that after a few revolutions the
satellite, losing progressively energy and angular momentum, enters the 
stellar 
distribution and suffers complete merger. This case illustrates  
the inadequacy of Chandrasekhar formula since it would  predict no drag
for a satellite orbiting outside the outer edges of the companion galaxy.
But not only, it clearly indicates that the satellite 
experiences a deceleration  that rises in response to the  
perturbation  
that is excited in the stellar field during the interaction.
This is a global phenomenon that can be explained in terms of an
interaction with resonant stars in the galaxy (Tremaine \& Weinberg 
1984). Thus a continuity  should exist 
in the physical  
processes inducing orbital decay.

The aim of this paper and of the accompanying one is to
clarify the role of the tidal and frictional forces 
 in affecting the motion of the satellite.
This is achieved within the context of the theory of 
{\it linear response}.
The method is close to that explored  by 
Bekenstein and Maoz (1992)   
who relate 
the frictional drag  to the stochastic  components of the background 
forces,  thus establishing a connection between dynamical friction
and the fluctuation-dissipation theorem.
Their method  proved useful in giving  a unified picture
of the process of energy loss and heating 
of a test particle traveling through a uniform stellar system.
In a subsequent work, Maoz (1993) provided  an approximate expression 
for the 
energy loss experienced by a massive particle moving  in a nonhomogeneous 
medium. He  noted that, owing to the nonuniformities present
in the underlying stellar system, dynamical friction
no longer depends on local background 
characteristics and that the  force acquires a component 
in the direction of the spatial density gradients.
%
Nevertheless he does 
not compute the force nor give a complete and consistent description of the 
process.

In this paper
we compute  the force of feed-back
on a satellite 
moving 
 in a background which is nonuniform and spherically symmetric,
improving 
the perturbative technique first developed by Kandrup (1981,1983;
see also the recent work by Nelson \& Tremaine 1997). 
The collisionless system of 
stars is 
described by the equilibrium distribution function $f_0.$ 
During the passage of the satellite $M,$ the stars
are subject to a time-dependent gravitational force 
and on account of that, 
the distribution
function modifies: Its new value $f=f_0+\Delta f$ is 
calculated using the theory of the linear response.
The perturbation induced by the satellite acts back producing 
a force $\vF_{\Delta}$ which is  computed, accordingly,  
as expectation value on the response  function $\Delta f$ 
of the microphysical
force resulting from the interaction of $M$ with the
stars of the background.
The characteristics  of this force are  closely related to the way the
stellar  background responds to the perturbation and depend, as  in 
memory effect, 
on the
whole history of the particle and stars 
since the moment in which the interaction is turned on.
The formalism also applies to the case of a satellite orbiting
outside of the stellar distribution.

In this paper we wish to shed light into the nature of the force and derive
a natural decomposition in terms of a global and local response
of the background.
This work is also preparatory 
to the analysis that will be  carried in paper II, on the role of tides in 
causing the
orbital decay of a satellite accreting  onto a companion galaxy, a process 
relevant  to the clustering of structures in cosmology.

In addressing all these problems,
for ease of analysis, and consistently with the hypothesis
that stars in equilibrium can be considered as a collisionless system, the
equilibrium distribution function $f_0$ is factorized.
This assumption is customarily used 
to compute mean properties of the response, as in Weinberg approach (1986).
We here find that the introduction of such hypothesis 
cancels out artificially  an important correlation present among the stars:
Under the action of their self-gravity, the stars  
perform bound orbits around the 
barycenter of the galaxy. The displacement that the 
center of mass  suffers during  the encounter with 
the satellite 
(owing to linear momentum conservation) 
represents a collective response neglected in previous analysis.
This response
needs 
to be
included for a correct estimate of the transfer of orbital energy into the
internal degrees of freedom of the galaxy.

In this paper we modify the formalism of the theory
of linear response to account for this
important effect.
This modification explains   the longstanding  problem
of the differences in the sinking times found 
in N-body simulations (White 1983; Zaritsky \& White 1988;
Hernquist \& Weinberg 1988) and in 
semi-restricted methods (Lin \& Tremaine 1983).
We present here the method used to include this effect but defer to 
paper II for a more thorough  analysis.

The layout of the paper is as follows: In $\S 2$ we sketch 
the method
used to compute the  macroscopic force on $M,$ 
using a derivation which is independent of that given by Kandrup (1981).
In $\S 3$ we test the theory against  known  results, focusing on shortlived
distant encounters.
In $\S 4$ we
derive the formal  expression of the force $\vF_{\Delta}$  on $M$ 
that accounts for the shift of the center of mass of the
galaxy.
We then compute in $\S 5$, the 
force on the satellite  $M$ moving at high speed
through a nonuniform stellar background 
characterized at equilibrium by 
a maxwellian distribution function.
In $\S 6$ we compute the frictional force  
on a satellite moving across a finite size cloud
with given density profile, and compute  
the shape  of the tidal distortion.
In $\S 7$ we evaluate the extent of the total energy loss.
In $\S 8$ we present our conclusions.
\vskip 30pt
\centerline {II. THEORY OF LINEAR RESPONSE}
\vskip 25pt
\centerline {2.1{\it Response Function and Back-Reaction Force}}

\vskip 20pt
Consider a particle (i.e., a satellite) of mass $M$ 
and velocity $\vecV$  moving through a
collisionless system of  
$N$ stars.
The stars have  mass $m\ll M$ and 
are described by an equilibrium distribution function
$f_0.$  
At $\vecR(t),$ the massive particle  experiences an $N$-body force 
resulting from the gravitational interaction with the background stars 
$$\vF(t)=-GMm\sum_{i=1}^{N}
{\vecR(t)-\vri\over \vert \vecR(t)-\vri\vert^3}\eqno (1)$$
where  $\vri$ is the position of the $i-$th star at time $t$.
By action-reaction, each of the $N$ stars of the background 
experiences a {\it perturbing} force 
$$\vF_i(t)=GMm
{\vecR(t)-\vri\over \vert \vecR(t)-\vri\vert^3}.\eqno (2)$$
Under the influence of the time dependent force $\vF_i,$ 
the stellar distribution function $f$
is altered
and  the  distortion with respect to equilibrium  can be described,
to first order in the perturbation, 
in terms of the {\it response} 
function $\Delta f$ such that $$ f= f_0 + \Delta f.\eqno (3)$$
 Because of the change 
in the distribution function induced by the satellite $M$, 
the force  on $M$ changes.
This 
{\it macroscopic} force can be  determined as
mean on the distribution function $f$ and is  decomposed into two terms 
 
$$<\vF>_{f_0+\Delta f}=<\vF>_{f_0}+<\vF>_{\Delta f}.\eqno (4)$$
The force  $<\vF>_{f_0}$  originates from the interaction of $M$
with the {\it smooth potential} generated by the {\it unperturbed} 
stellar system; the term $<\vF>_{\Delta f}$ instead will represent 
the force of {\it back-reaction} whose magnitude  depends
on  the response function $\Delta f$.
As a first
step, we calculate the perturbation $\Delta f$ on the stellar
distribution function.

Let $-\vF=\sum_{i=1}^N\vF_i$ be the applied time dependent force 
acting on the stellar background.
The distribution function $f=f_0+\Delta f$
can be computed,
to first order in the perturbation, 
 using the theory of {\it linear response} (TLR hereafter).
For this purpose we  need to specify the  form of the disturbance and the 
properties of the stellar system in the {\it unperturbed} state
through the equilibrium  Hamiltonian $H_0.$

In the 
collisionless stellar system
$$H_0=\sum_{i=1}^{N} {p_i^2\over 2m}\,\,\,\,+ \Psi.\eqno (5)$$
where $\Psi$ describes 
the smooth self-interaction, i.e. the mean field
of the gravitating system.
In presence of the 
time dependent external force induced by the satellite, the Hamiltonian  
modifies 
and the perturbation $H_1$ 
is simply  
$$
H_1(t)=
-\sum_{i=1}^N{GMm\over \vert\vecR(t)-\vri\vert}. 
\eqno (6) $$
\noindent
$H_1$ results from the superposition of the 
gravitational force that the incoming
body exerts on each particle of the system. 
$H_1$ is regarded as an explicit function of $t$ depending 
on the instantaneous position $\vecR(t)$ of the 
particle $M$ and
as a  function of the 6$N$ coordinates ($\vec r$, $\vec p$) 
of the phase space: hence, 
$H_1=H_1(t;\Gamma).$  ($\Gamma$ denotes shortly
the  phase-space coordinates)

According to the  Liouville equation, 
the evolution of the distribution function $f$ is given by the equation 
$${\partial f(t)\over \partial t}=\{H_0+H_1(t),f(t)\}.\eqno (7)$$
If the distribution function $f_0$ 
depends on the phase-space coordinate only through $H_0$ 
the 
Poisson bracket $\{ H_0,f_0\}$  
vanishes identically, and   
to first order in the perturbation, we can approximate equation (7) as

$${\partial \Delta f(t)\over \partial t}=\{H_0,\Delta f\}+
\{H_1(t),f_0\}.\eqno (8)$$ 
The formal solution is
$$\Delta f(t;\Gamma)=\int _{-\infty} ^{t}\,ds\,\, {\rm 
{e}}^{-i(t-s){\Liu}}\,\, \{H_1(s),f_0\},\eqno (9)$$
where ${\Liu}\equiv i\{H_0,\,\,\}$ denotes the Liouville
operator acting on the $6N$ coordinates of the phase space.
In equation (9) the expression of the  Poisson bracket reads 
$$\{H_1(s;\Gamma),f_0\}=
-GMm\sum_{i=1}^N{
\vecR(s)-\vr_i\over 
\vert \vecR(s)-\vri\vert ^3}
\cdot \grad_{\vpi} f_0.\eqno (10)$$


In defining the limits of integration of equation (9), 
the external force is  applied in the infinite  past when the 
system was
in a stationary state and the particle $M$
far away: accordingly, $H_1$ vanishes for $t\to -\infty$.

Having calculated the distortion  $\Delta f,$ (to first order in the
perturbation $H_1$) we  can compute the expectation value  
of the microphysical force $\vF$ acting on the satellite $M$ at time $t:$
$$<\vF(t)>_{\Delta f}\equiv \int\,\, d\Gamma\,\, \Delta f\,\, \vF(t;\Gamma).
\eqno (11)$$
For a perturbation of the form given by equation  (10) 
$$<\vF(t)>_{\Delta f}=- GMm\sum_{i=1}^N\,\, \int_{-\infty}^{t}ds
\int d\Gamma\,\,e^{is\Liu}\left [\nabla_{\vpi}f_0\cdot 
{\vecR(s)-\vri\over \vert \vecR(s)-\vri\vert^3}\right ]
\,\,e^{it\Liu}\left [\vF(t;\Gamma)\right ]\eqno(12)$$
where we used the  property that $\Liu$ 
is a self-adjoint operator. 
In equation (12) the Liouville operator acts on the
phase-space  coordinates 
($\Gamma$): applied to a generic dynamical variable
 $A$ (representing any component of
the phase-space coordinates which satisfies  equation  
${\partial A/ \partial t}=i{\Liu}A$ )
$\Liu$ carries $A$ along its evolution from time $t=0$ to time $t$ , i.e., 

$$A(\Gamma(t))={\rm {e}}^{i t {\Liu}}A(\Gamma (0))\eqno (13)$$
Hence, we interpret 
${\rm {e}}^{it {\Liu}}\vF$ at an arbitrary time $t$ 
as the value of 
the force $\vF$ acting on $M$ at $\vecR(t)$
resulting from the gravitational interaction with the particles
that evolved from $t=0$ to time $t$ along a
path whose dynamics is uniquely determined
by the unperturbed  Hamiltonian $H_0$. Likewise, $e^{is\Liu}$ acts
on the scalar product to yield accordingly

$$
<\vF(t)>_{\Delta f}=
-GMm\,\sum_{i=1}^N\int_{-\infty}^{t}ds\,\,
\int \,\,d\Gamma \,\,\left [ \grad_{\vpi(s)} f_0\,\, \cdot 
{\vecR(s)-\vri(s)\over \vert \vecR(s)-\vri(s)\vert^3}\right ] 
\,\,\vF(t;\Gamma(t))\eqno (14)$$
We will shortly denote $<\vF(t)>_{\Delta f}=\vF_{\Delta}$ hereafter.
This equation  provides the formal expression of the force
acting on $M$ in terms of  the microscopic force $\vF$ 
and of the equilibrium distribution function $f_0$.  
We notice that within TLR, the response function
$\Delta f$  is of order ``$G$''; Hence, the resulting 
force describing the feed back mechanism acting on $M$ 
is of order ``$\,G^2\,$''. 
Equation (14) contains all details of the 
microphysical interaction 
of the $N$ stars with the perturber:  
It accounts for the {\it self-gravity} of the stellar equilibrium 
and for the actual dynamics of the perturber.

The difficulty in evaluating the multiple integrals of equation (14), 
involving  the actual dynamics of the underlying stars, is enormous particularly
if we want to use analytic tools to gain insight on the mechanism
producing friction.
To overcome this problem we 
introduce a major simplification knowing that  
the stellar motion in a galaxy can be described in terms of  
a mean field  potential (Binney \& Tremaine 1987): 
The system can therefore  be regarded as  collisionless and stars behave as
independent particles.
The corresponding distribution function $f_0(\Gamma)$
can therefore be written in terms of the one-particle
phase space density  $\fop$:
$$f_0=\Pi_{i=1}^{N}{\fop(\vr,\vp)}.\eqno (15)$$

Under this hypothesis, due to the statistical independence of the particles, 
the cross-correlation terms (i.e., those involving
different indices in the summation of eq.[14]), cancel 
identically,  in the limit $N\gg 1$.
Only the term of self-correlation survives
giving rise to a macroscopic force 
$$\vF_{\Delta }
=[GM]^2Nm^2\int_{-\infty}^t ds
\int d_3\vr \,d_3\vp \,\,\left [ 
{\grad_{\vp(s)}}\,\, \fop\cdot  {\vecR(s)-\vr(s)\over
\vert \vecR(s)-\vr(s)\vert ^3}\right ] \,\,
{\vecR(t)-\vr(t)\over
\vert \vecR(t)-\vr(t)\vert ^3}\eqno (16a)$$ 
where the phase-space volume $d_3\vr\,d_3\vp$ and $\fop$ 
are independent of time and can be referred to time $t.$
An equivalent expression for the force can be obtained by applying 
the evolution operator $e^{-i(t-s)\Liu}$ to $\vF(t;\Gamma)$ in equation (12)
and reads 
$$\vF_{\Delta }
=[GM]^2Nm^2\int_{-\infty}^t ds
\int d_3\vr \,d_3\vp \,\,\left [ 
{\grad_{\vp}}\,\, \fop\cdot  {\vecR(s)-\vr\over
\vert \vecR(s)-\vr\vert ^3}\right ] \,\,
{\vecR(t)-\vr(t-s)\over
\vert \vecR(t)-\vr(t-s)\vert ^3}\eqno (16b)$$ 
(equation (16b) will be used in paper II).

Introducing the velocity vector 
$\vv=\vp/m,$ we can express $\fop$ as a function of $(\vr,\vv)$.
If the equilibrium system is described by
 a Gaussian distribution 
function $$\fop(\vr,\vv)=n_0(r)\left ( {\beta m\over 2 \pi}\right )^{3/2}
\exp(-{1\over 2}\beta m v^2)\eqno (17)$$
we have $\grad_{\vv}=-\beta \fop\,\,\vv$.
In equation (17),  $r$ and $v$ denote the absolute values of $\vr$ and $\vv$
and the coefficient $\beta\equiv(m\sigma^2)^{-1}$ is directly 
related to the one-dimensional
dispersion velocity $\sigma$ characteristic of the 
equilibrium stellar distribution;
$n_0(r)$ is the unperturbed probability density which 
for a spherically symmetric
system is a function of $r$ only. $\fop$  is defined so that
$\int d_3\vr \,\,n_0(r)=1.$
With the change of variables $(\vr,\vp)\to(\vr,\vv)$ and according to the 
normalization used we recall that 
$d_3\vr\,d_3\vp\fop(\vr,\vp)$
is equal to 
$d_3\vr\,d_3\vv\fop(\vr,\vv).$

%
%

The force $\vF_{\Delta}$  at time $t$ is found to  depend 
 on the whole  `` history "  of the composite system, i.e.,  
 on the dynamics of the satellite, and on the dynamics 
of stars of the  background as 
determined 
by the (mean field) Hamiltonian $H_0$.
If we introduce the tensor 
$$T^{ba}\equiv -[GM]^2Nm^2\beta
{R^b(s)-r^b(s)\over \vert \vecR(s)-\vr(s)\vert^3}
{R^a(t)-r^a(t)\over \vert \vecR(t)-\vr(t)\vert^3}\eqno (18)$$
correlating the microscopical force at time $s$ with that at time $t,$
we can interpret equation (16) written in the form  
$$F^a_{\Delta}=\int_{-\infty}^{t} \,ds \,\, \int d_3\vr\,\,d_3\vv \,\,
\fop v^b(s)T^{ba}\eqno (19)$$
as 
a manifestation on the {\it fluctuation-dissipation theorem} relating the 
averaged
force of back-reaction to the time integral of a suitable dynamical
correlation function of the unperturbed system (Bekenstein \& Maoz 1992;
Nelson \& Tremaine 1997).
Due to the long range nature of the gravitational interaction
all stars in the galaxy give a sizable contribution to the force. 
$\vF_{\Delta}$ is a   force
which is intrinsically 
nonlocal both in phase-space and time.
In equation (19), $v^b\,T^{ba}$ are shorthand for the summations (hereafter).

Notice that in the steps that lead to  equation (19)  
the perturber's motion  needs not necessarely to be confined within 
the stellar distribution. 
We can therefore  evaluate, within TLR, the force
on the  massive particle $M$ when it is orbiting outside the galaxy.

\vskip 25pt
\centerline {2.2 {\it Force and Density Distortion}}
\vskip 20pt

$\vF_{\Delta}$ can be alternatively  viewed 
as a force rising in response to the distortion caused by the satellite
on the  stellar density field. 
The satellite $M$ induces a disturbance in the
density distribution function $n(\vr)$ 
that can be readily evaluated, within TLR, from equation (9).
Relative
to its unperturbed value $n_0$, the density changes by an amount equal to
$$\Delta n(\vr)=\int d_3\vr''\,\,d_3\vv \,\,\delta_3(\vr''-\vr)
\Delta f(\vr'',\vv)\eqno (20)$$
where $\delta_3$ denotes the 3-dimensional Dirac function.
From the solution (9) and equation (10) we have

$$\Delta n(\vr)=-GM\int_{-\infty}^{t}\, ds\,\,\int d_3\vr''\,d_3\vv
\delta_3(\vr''(t-s)-\vr)\grad_{\vv} \fop (\vr'',\vv)
\cdot {\vecR(s)-\vr''\over \vert \vecR(s)-\vr''\vert^3}.\eqno (21)$$
In the steps that lead to equation (21) we have used explicitly the
self-adjointness 
property of the Liouville operator and the factorization of $f_0$
(
$\cal {L}$ is applied to the phase-space coordinates of the 
Dirac function). 
Under the  disturbance (21), the satellite experiences a force 

$$\vF_{\Delta}=-GMNm\int d_3\vr \Delta n(\vr){\vecR(t)-
\vr\over \vert \vecR(t)-\vr \vert^3}\eqno (22)$$ 
which coincides with equation (16b).
The  expression of 
the force $\vF_{\Delta}$ will  be modified in $\S 4$ to account properly
for the shift of the centre of mass of the stellar distribution,
that occurs during the  gravitational interaction.

Given $\vF_{\Delta}$, we can formally compute the total energy loss
suffered by the satellite
$$\Delta E=\int_{-\infty}^{+\infty}dt\,\, \vF_{\Delta}\cdot \vecV=
-{1\over 2}[GMm]^2N\beta \int d_3\vr\, d_3\vv\,\,\fop
{\cal {J}}^2\eqno (23)$$ 
where
$${\cal {J}}=\int_{-\infty}^{+\infty}dt \,\, \vv \cdot {\vecR-\vr\over
\vert \vecR-\vr\vert^3}.\eqno (24)$$
Hereafter $\vecR$, $\vr$, and $\vv$  will denote the vectors  
at the current time $t$ and the volume element $d_3\vr\,d_3\vv$
can be  referred only for convenience to time $t.$
(The derivation of equation (23) is summarized in Appendix A.)

\vskip 30pt
\centerline {III. TESTING THE THEORY}
\vskip 25pt

Kandrup (1983) verified, in his early work, that
equation (16) reproduces the expression
of the drag force derived by Chandrasekhar for  the case of  a point
mass moving through an infinite homogeneous 
stellar background.
This important limit is recovered  within TLR 
in the hypothesis that the stars and the perturber move 
along straight lines (we do not sketch here the derivation
of  this limit but refer to Kandrup). 
According to the original formulation (Chandrasekhar 1943),
the 
gravitational interaction among the  stars was neglected and  
the braking force  on $M$ was derived considering the momentum exchange 
resulting from the incoherent 
superposition of the binary encounters  
with single stars moving along hyperbolic  orbits relative to $M$.
TLR reproduces the correct results to
order $G^2$, giving in the high speed limit, i.e, for $V\gg \sigma$
 a frictional force 
$$ 
\vF_{DF}=-4\pi {[GM]^2\over V^2} Nmn_0\ln \left ({b_{max}\over 
b_{min}}\right ) {\vecV\over V}\eqno (25)$$
where $b_{max}$ is set 
equal to the characteristic length of the stellar system 
to avoid a ``infrared " divergence. 
In TLR, the above expression is derived introducing also 
at short distances
a ``ultraviolet'' cutoff, $b_{min}$, the minimum impact parameter
for collisions at large scattering angle.
The presence of a  divergence
at small relative distances between $M$ and the stars, 
in our formulation, is a consequence of the
approach used. 
Since we have treated $H_1$ as a small perturbation, equation (9)
becomes invalid when the distortion in the background
produces a wake with overdensity
$\Delta n/n_0\sim 1$.
The density enhancement $\Delta n/n_0$ 
(computed using equation (9) and neglecting the self-gravity 
of stars)  
$${\Delta n \over n_0}={GM\over \sigma^2}{1\over
\vert \vecR-\vr\vert}e^{(\alpha^2-x^2)}\,[{\rm {erf}}\alpha +1]
\eqno (26) $$
(with $\alpha\equiv (\vecV/\sqrt {2}\sigma)\cdot (\vecR-\vr)/\vert
\vecR-\vr\vert,$ and $x\equiv V/\sqrt {2}\sigma$) 
becomes
close to unity when the relative distance between
a star and the satellite $M$ is 

$$\epsilon\sim {GM\over {\rm {Max}}(\sigma^2,V^2)}\eqno (27)$$

%
One can recognize that at such a relative distance the mean squared velocity 
of a typical
field particle would increase by an amount of order $\sigma$ ($V,$ in the 
high speed limit). 
For an extended  (nondeformable) satellite 
$\epsilon$ is instead set equal to
physical extension of $M.$  Thus $\epsilon$ gives the 
``permitted" size the satellite (White 1976).
In the expression of the force $\vF_{\Delta}$ the domain of integration 
is thus limited to $\vert \vecR-\vr\vert\gg\epsilon$.

Can the theory of LR
be tested against other known results  ?
The energy loss of an object of mass $M$  moving at high speed $V$ 
relative to a galaxy  characterized by a  mean-square
radius $<r^2>$ is known to be
$$\Delta E= -{4\over 3}Nm<r^2>{[GM]^2\over
V^2\,\,b^4}. \eqno (28)$$
This equation is derived in  the hypothesis that 
the distance of closest approach
 $b$ largely  exceeds the radius of the disturbed galaxy  
and that the 
effective duration of the encounter
$\tau_e$ be shorter than the crossing (dynamical) time
of the stars  $\tau_{dyn}\sim (<r^2>/<v^2>)^{1/2}$
where $<v^2>$ 
is the mean squared velocity ($<v^2>=3\sigma^2$).
This limit is known as impulse approximation and applies
when the perturber moves at high speed ($V\gg \sigma$)
 relative to the
stars.
The energy loss (28) is of order ``$G^2$''and was derived by Spitzer
(1958)  on the basis of simple arguments,  without direct knowledge
of the force causing such a loss (Binney \& tremaine 1987).

Using equation (23)    we recover Spitzer's results as 
in high speed encounters 
the energy loss of the satellite is exceedingly small (scaling 
with $V^{-2}$), and  TLR applies.  
The
motion of  the perturber can 
be approximated as linear while 
stars barely move
since the interaction  is shortlived.
Under these simplifications, equation (23) for the energy loss
of $M$ can be evaluated analytically 
(the calculation is sketched in Appendix A) to give
$$\Delta E=-2Nm{[GM]^2\over V^2\,\,b^2}
-{4\over 3}Nm<r^2>{[GM]^2\over 
V^2\,\,b^4}+O\left ({<r^2>\over b^2}\right)\eqno (29)$$
$\Delta E$ as expected is 
the sum of two contributions. The first 
term, according to linear momentum conservation,
 is due to {\it recoil}, i.e.,  the effect representing the exchange of energy
between $M$ and the galaxy viewed as  pointlike.
The second instead describes the transfer of orbital energy
into the internal energy of   
the stellar distribution. 

\vskip 30pt
\centerline { IV. TLR REVISED:
THE FORCE IN THE GALAXY REST FRAME}
\vskip 25pt
\centerline {4.1 {\it  Correlation and shift of 
stellar barycenter}}
\vskip 20pt
In this paper, we wish to  gain insight into the
mechanisms that ultimately cause 
the transfer of orbital energy into the internal degrees of freedom
of the stellar distribution.
For this purpose, 
we need to  isolate from the force $\vF_{\Delta}$ those terms
responsible to the recoil, representing 
only a  global displacement  of the collection of stars
in the system with  no relation to the 
"binding" mechanism that $M$  experiences as a result of friction.
We note that
equation (16) derived
under the hypothesis that stars, in equilibrium, behave as
{\it  
independent 
particles}, does not correctly describe 
the energy transfer during   the interaction.
At the basis of this inconsistency is 
the simplifying assumption of factorization of $f$,
the equilibrium distribution function (eq. [15]).
According to this 
hypothesis, the dynamics of each star is 
uncorrelated from the dynamics   of any other star in the galaxy.
In this approximation each star moves in the averaged  potential
generated by all the other stars.
For a spherical distribution this potential gives rise to a central 
force binding the stars to the center of mass of the galaxy.
In the encounter 
with a satellite, the center of symmetry of the stellar
distribution suffers  a rigid 
displacement 
due to linear momentum conservation. 
The stars thus coherently move responding to the satellite
perturbation with a coordinated  shift of their orbits.   However, 
this correlation  (present in the full formalism leading to eq. [14]) is 
lost, since the mean field dynamics we use to 
represent the unperturbed  motion of the stars (eq. [15]) does not allow for 
the shift  of the center of mass; the dynamics  has 
lost its {\it translational invariance}.

As a consequence, equation (23) for $\Delta E$ 
 gives a consistent estimate of
the energy loss
only in three limiting 
cases;  when (i) 
the satellite moves 
through a homogeneous infinite background of non-interacting  stars, 
(ii) when the interaction is shortlived, as 
in the impulse approximation or (iii)   when the center of symmetry
of the galaxy is {\it pinned}.
In the encounter of a satellite with a spherical
galaxy this last hypothesis is clearly
unphysical.
The above considerations  rise therefore 
a problem of consistency which can not be overlooked.

To solve for  this difficulty,
we examine the process of dynamical friction not from 
the laboratory system
but from the
noninertial frame of reference
comoving with the center of mass of the perturbed galaxy:
$\vecX=(1/N)\,\sum_i\vr_i$.
Newton's equations, in this frame, read
$$\mu {d^2(\vecR-\vecX)\over dt^2}=<\vF>_{\fop}+ \vF_{\Delta}\eqno (30)$$

$$m {d^2({\vr_i-\vecX})\over dt^2}=\vF_{internal} + (-\vF_i) - {1\over N}
\sum_{j=1}^N (-\vF_j).\eqno (31)$$
The satellite simply acquires a new mass equal to the reduced
mass of the system $\mu=M Nm/(M+Nm).$ 
Instead, the rhs of equation (31) for the motion
of the $i-$th  star acquires a new term, representing
the opposite  of the ``mean" force exerted by the satellite on the galaxy,
as a whole ($\vF_i$ is the force
acting on $M$ due to the $i-$th star; eq. [2]).

Guided  by equations (30) and (31), we are led to modify  $\vF_{\Delta}$  
introducing a new correlation tensor $T$ obtained subtracting
such a `` mean" force to the kernel
$[\vecR-\vr]/\vert \vecR-\vr\vert^3$ along the complete history until time $t$,
yielding

$$
\eqalignno{
\vF_{\Delta}= &
-[GM]^2Nm^2\beta\int_{-\infty}^t ds
\int d_3\vr\,d_3\vv 
\, \fop
\cr &
\left \{ \vv(s)\cdot 
\left [
 {\vecR(s)-\vr(s)\over
\vert \vecR(s)-\vr(s)\vert ^3}
-\int d_3\vr' n_0(r'){\vecR(s)-\vr'\over \vert \vecR(s)-
\vr'\vert^3}\right ]
\right \}
\,
{\vecR-\vr\over
\vert \vecR-\vr\vert ^3}.& (32)\cr}$$
$\vF_{\Delta}$ represents here the response  force acting of $M$ 
at time $t$ as measured
in the frame comoving with the perturbed stellar distribution.
We now regard $\vecR$ and $\vr$ as coordinates relative
to the center of mass of the perturbed stellar system.

In this frame, the disturbance
in the density field is calculated according to 
equation (32), by introducing 
the modified kernel. For a Gaussian distribution function,
we have 

$$
\eqalignno{
\Delta n(\vr)=
&
GMm\beta \left({m\beta\over 2\pi}\right )^{3/2}
 \int ds\,\int d_3\vr''\,d_3\vv
\,\, \delta_3(\vr''(t-s)-\vr)n_0(r'')e^{- [ {1\over 2}\beta mv^2]}
\cr  & 
\vv\cdot \left [{\vecR(s)-\vr''\over \vert \vecR(s)-\vr''\vert^3}
-\int d_3\vr'\,\,n_0(r'){\vecR(s)-\vr'\over \vert \vecR(s)-\vr'\vert^3}
\right ]&  (33)  \cr}$$
with $\vr'$ a dummy variable. 
Equation (32) (or equivalently eqs. [33] and [22]) 
provides the {\it correct} expression for the force $\vF_{\Delta}$ 
on 
the  satellite moving in the gravitational field of  the primary galaxy.

\vskip 25pt
\centerline {4.2 {\it Self-gravity of the response}}
\vskip 20pt
The shift of the stellar barycenter was viewed by Hernquist and Weinberg
(1989) as a manifestation of the self-gravity of the response.
According to our  analysis, 
the modified kernel in equation (32) derived to account   for this 
displacement is introduced  to restore  the correlation
cancelled in the formalism of TLR when $\fop$ is introduced to represent
the equilibrium stellar system.
This is  a key correction that accounts  for the 
{\it self-gravity
of the stellar background.}
We underline the fact that in TLR the distribution function
is perturbed only by the satellite {\it external}  potential.  
The {\it self gravity of the response} has a different
origin (White 1983; Nelson \& Tremaine 1997): It expresses
 the ``change" in 
the galaxy's self-interaction potential induced by 
the perturbed  density field (excited by the intruder). 
This would result in an additional component to the force $\vF_{\Delta}$
  on the satellite which is of higher order, and 
for this reason is neglected in our scheme.
In fact, the response function $\Delta f$ modified to account for the 
perturbed self-gravity would 
be of order $G^2$, leading to correction of order $G^3$ on the 
drag force. 

This considerations are  consistent with  results, from numerical 
simulations
by Bontekoe \& Van Albada (1987)  and Zaritsky \& White (1988), 
on the negligible role
played by the perturbed self-gravity in affecting the process of energy and
angular momentum transfer by the satellite. The self-gravity of the 
equilibrium system is included and affects the process of energy transfer
(as it will be shown in \S6.5).

\vskip 30pt
\centerline {V. FRICTION AND TIDAL FORCE DECOMPOSITION}
\vskip 25pt
\centerline {5.1 {\it  High speed encounters}}
\vskip 20pt
In this Section we examine the case of a satellite $M$ 
moving at high speed relative to a spherically symmetric stellar background.
 In this limiting case  and to order ``$G^2$'', 
the dynamics of the satellite and that of the stars can be approximated as
 $$\vr(s)=\vr-(t-s)\vv\eqno (34)$$
 $$\vecR(s)=\vecR-(t-s)\vecV,\eqno (35)$$
with $\vecV$ and $\vv$ constant vectors. The satellite and the stars 
thus move along straight lines, and we neglect 
the self-gravity of the stellar equilibrium system.
The validity of this approximation thus restricts to
encounters whose duration does not exceed
the crossing time of the stars in the galaxy.  
In the high speed limit, 
the density disturbance thus reads

$$
\eqalignno{
\Delta n(\vr)=
&
GMm\beta \left({m\beta\over 2\pi}\right )^{3/2}
 \int_{-\infty}^{t} ds\,\int d_3\vv
\,\, n_0(\vert \vr-(t-s)\vv\vert )e^{ [-{1\over 2}\beta mv^2]}
\cr  &
\vv\cdot \left [{\vecR-(t-s)\vecV-\vr+(t-s)\vv\over \vert 
\vecR-(t-s)\vecV-\vr+(t-s)\vv \vert^3}
-\int d_3\vr'\,\,n_0(r'){\vecR-(t-s)\vecV-\vr'\over \vert \vecR
-(t-s)\vecV-\vr'\vert^3}
\right ] &  (36)\cr }$$
as the presence of the Dirac function
$\delta_3(\vr''(t-s)-\vr)$  in the coordinate space
relates the density at $\vr$ to  the density at $[\vr-(t-s)\vv].$
The spherical symmetry is lost since the satellite produces
a distortion that keeps memory of the
star and satellite motions; as a consequence, $\Delta n$ is a function
of vector $\vr.$

The density response, when $V\gg \sigma,$  should not depend on $\sigma$
explicitly and its expression can be estimated in the limit
$\beta\to\infty$. 
The relevant contributions to $\Delta n$
arise, in this limit, from
those stars having very low velocities for which $\beta v^2$ is finite.
Hence, to $O(1/V^2),$ the density response $\Delta n$
can be evaluated expanding the arguments of the integrals 
to first order in the velocity $\vv$ (the calculation is 
sketched in Appendix B). 

We find that the response function $\Delta n(\vr)$ 
naturally decomposes  into two  contributions: the first
representing the wake rised by the satellite
along his trail: The stream of particles
behind $M$ is characterized by an overdensity
$$
\Delta_1 n(\vr)=4\pi GMn_0(r)\int_0^\infty  d\tau\,\,\tau
\delta_3(\vecR-\vr-\tau\vecV).
\eqno (37)
$$
The second contribution rises from the density
inhomogeneities of the underlying unperturbed stellar system.
Introducing 
the unit vector
$\vecn\equiv \vecV/V,$  we have

$$\Delta_2 n(\vr)=-{GM\over V^2} \vecd_{(\vecR-\vr)}\cdot 
\grad_{\vr} n_0(r)\eqno (38)$$
with 
$$\vecd_\vecY=\vecs+\vecn\left [\ln{
(\vert \vecY  \vert)-\vecY \cdot \vecn\over R-\vecR\cdot\vecn}
\right ]
+{\vecY-\vecn (\vecY\cdot \vecn)\over 
(\vert \vecY \vert )-\vecY\cdot\vecn}
-{\vecR-\vecn(\vecR\cdot\vecn)\over
R-\vecR\cdot\vecn}.\eqno (39)$$
The distortion vector $\vecd$ gives the description of the global tides
excited by the satellite during the encounter.
In equation (39) the  vector $\vecs$ 
is defined as 
$$\vecs(\vecR)=\int_0^{+\infty} \, dx \,
x {\vecR-\vecn x \over \vert \vecR-\vecn x \vert^3}
\lb(\vert \vecR-\vecn x \vert )\eqno (40)$$   and 
$$\lb(r)\equiv \int_{r'>r}\,d_3\vr'n_0(r')\eqno (41)$$
We have verified that the center of mass of the perturbed
stellar system is not affected by the 
density distortion $\Delta n=\Delta_1 n
+\Delta_2 n$. The displacement vector $\vecs$ compensates for the shift
of the center of mass caused by the global tide, in the
frame comoving with the galaxy.
The result illustrates the importance of defining the 
appropriate physical frame of reference for the description of the
gravitational interaction between the satellite and the primary stellar 
system.

Given the density distortion $\Delta n$, we can calculate the 
force acting on the satellite.
We find that  $\vF_{\Delta}$ results in the
superposition of two components.,  $\,\,\vF_{\Delta}=\vF_{DF}+\vF_T$,
the first
representing the frictional drag induced by the density
enhancement  $\Delta _1 n$  
that originates behind $M$ 
$$\vF_{DF}=-\vecn\,\,
4\pi{[GM]^2\over V^2}\int_{\epsilon/V}^{\infty} 
d\tau {\rho_0(\vert \vecR-\tau \vecV\vert )\over \tau}
\eqno (42)$$
where the stellar mass density is defined as $\rho_0(r)=N m n_0(r)$.
In the expression of the force, the Coulomb logarithm 
(eq. [25])  is  replaced by 
an integral, i.e., an "effective" $\ln \Lambda$ 
which 
depends  on the density distribution along the wake, since the vector
$\vecR-\tau \vecV$ selects only those stars streaming behind the satellite
in its motion. We thus find that {\it dynamical friction rises
from a memory effect involving only those
stars perturbed along the motion of the satellite}.
We may consider, loosely speaking,  this contribution to the force as
{\it local}
since the drag on $M$
is induced by the bow shock that the satellite excites 
in its supersonic motion across the medium (see eq. [26]).
Equation (42) reproduces Chandrasekhar formula for an infinite homogeneous 
medium.

In addition to the frictional force, we discovered  a new component 
that  is closely related
to the {\it tidal distortion} induced by the satellite in its passage.
To first order  in the perturbation, 
this component  is
$$\vF_{T}={[GM]^2\over V^2}\int d_3\vr\,\, \vecd_{(\vecR-\vr)}\cdot
\grad_{\vr} 
\rho_0(r) \,\,
{\vecR-\vr\over \vert\vecR-\vr\vert^3}.\eqno (43)$$
The force $\vF_{T}$ involves a volume integral (contrary to
$\vF_{DF}$) depending on the density gradients of the unperturbed 
stellar system: It therefore accounts for the {\it nonlocal}
response of the galaxy to the disturbance and can be viewed 
as a force resulting from the tidal deformation 
rised by the satellite  during the gravitational interaction.
We will refer to as {\it tidal force} thereafter.
Because of the nonhomogeneity  of the underlying stellar background,
$\vF_T$ is a force acting not exclusively along the direction of motion:
In a spherical system it  acquires a component along $\vecR$, i.e., 
along the vector joining the center of the galaxy to the instantaneous 
position of the satellite.  Notice that the 
decomposition of the force 
in terms of a tidal and frictional drag 
has been possible since  we have described the
 interaction of the spherical system 
 from the noninertial frame of reference 
comoving with the center of mass of the galaxy:
The galaxy, as viewed in 
the laboratory frame, suffers in addition the recoil.

\vskip 30pt
\centerline {VI. DYNAMICAL FRICTION IN FINITE SIZE CLOUDS}

\vskip 25pt
\centerline {6.1. {\it Uniform clouds}}
\vskip 20pt

As a first application, we estimate the frictional force experienced
by a satellite traveling through a homogeneous spherical 
cloud of radius $L$. Consistently with equation (35), the
satellite motion
is approximated as uniform and introducing the impact parameter 
$\vb$ defined so that 
 $\vb\cdot\vecV=0$ we have 
$$\vecR(t)=\vb+t\vecV.\eqno (44)$$
If $\vecV$ lies  along the $x$-axis, and $\psi$ denotes the
angle between $\vecV$ and $\vecR$ (varying in between $(0,\pi)$), 
the satellite is seen to  enter (exit)  the  stellar distribution
at an angle $\psi_1$ ($\psi_2$) for which
$$\cot\psi_1= -{1\over b}(L^2-b^2)^{1/2}\eqno (45)$$
$$\cot\psi_2= +{1\over b}(L^2-b^2)^{1/2}\eqno (46)$$
In this  case, the logarithmic integral  that appears in equation (42) 
is readily estimated. When the satellite moves {\it inside} the cloud
(i.e., when $\psi_1<\psi<\psi_2$ )
it experiences a frictional force
$$\vF_{DF}=-\vecn\,\,
4\pi{[GM]^2\over V^2}
\rho_0\,\ln \Lambda_{\rm {in}}
\eqno (47)$$
where  
$$
\ln \Lambda_{\rm {in}}=\,\ln{b\cot \psi +(L^2-b^2)^{1/2}+\epsilon 
\over \epsilon}.
\eqno (48)$$
In Figure 1, the effective Coulomb 
logarithm  ($\ln\Lambda$; solid line) is plotted against  $\psi$ for two 
selected values  of $b/L$ 
and for a cusp $\epsilon/L=GM/LV^2=0.01$. 
 We notice that the frictional force, and in turn $\ln\Lambda$
increases in magnitude as the massive particle  progresses 
along its motion inside the cloud:
Early in the encounter  the force is small  but as the wake develops 
the drag rises. For sufficiently small (large) impact
parameters relative to $L$, 
the force stays constant (steadily increases) 
until the edge of the cloud is reached.
For $b\ll L$ the Coulomb logarithm becomes nearly 
independent of $b$ and $b_{max}\to L$, the size of the cloud.

At angles $\psi<\psi_2$ the satellite finds itself outside of the 
stellar distribution. Despite that, 
the drag force is non vanishing because  the  stream of particles excited 
during its passage still produces a deceleration. 
This is a consequence of the memory effect
present in the correlation tensor. The contribution to
the force 
is given by

$$\vF_{DF}
=-\vecn\,\,
4\pi{[GM]^2\over V^2}
\rho_0\,
\ln 
{b\cot \psi+(L^2-b^2)^{1/2}+\epsilon \over b\cot \psi -(L^2-b^2)^{1/2}
+\epsilon }
\eqno (49)$$
with $\psi \le \psi_2.$
For $\psi\to 0$, i.e., for $t\to\infty,$ the logarithm 
slowly vanishes as $(2/b)(L^2-b^2)^{1/2}\tan\psi\propto 1/t.$

In a real system, the force will decay  as soon as the stars 
return to equilibrium, 
after a time $t\sim 1/\omega,$
where with $\omega$ we denote the mean internal frequency 
($\omega=<v^2>^{1/2}/<r^2>^{1/2}$).
In the high speed limit, stars respond to the perturbation
induced by the satellite 
as dust  particles.  Hence equation (49)  is   valid 
as long as the stellar  dynamical time exceeds
the time scale of the encounter $\tau_e$, or equivalently 
${\omega R/ V}\ll 1.$

Notice that in addition to the frictional force $\vF_{DF},$
TLR predicts also a tidal contribution  that we will evaluate in $\S 6.3$.


\vskip 25pt
\centerline {6.2 {\it Nonuniform spherical galaxy}}
\vskip 20pt
In a spherical system with unperturbed density profile
$$
\rho_0(r)=\rho^*_0\left [1+\left ({r\over L}\right )^2\right]^{-\gamma}
\eqno (50)$$
the satellite approaches initially  regions 
of  increasing  density and consequently suffers a drag whose extent 
with time (i.e., with decreasing $\psi$).
At $\psi<\pi/2$ the medium begins to rarefy  and the
braking force weakens as the wake develops in a medium of decreasing 
density.

 The magnitude of the force as a function of the orbital phase
$\psi$ can be computed analytically from equation (42) for  $\gamma=3/2,$ 
defining 
the ``effective"  Coulomb
logarithm as 
$$\ln \Lambda=\int _{\epsilon/V}^{\infty}{d\tau\over \tau}
\left [1+\left ({\vert \vecR-\tau \,\vecV \vert\over L}\right )
^2\right ]^{-\gamma}.\eqno (51)$$  
In this model, 
the mass within  radius $r$ declines
logarithmically and the satellite in its motion remains always confined within
the star's background.
Figure 2 shows $\ln\Lambda$ for two values of the impact parameter
$b=0.9$ and $0.5,$ 
(respectively) expressed in units of $L,$ the radial scale below which the
density profile is almost uniform.
We find that if the satellite 
moves in the halo of declining density 
only grazing the uniform density core, 
the drag reduces ( by a factor of $\sim2$ for $b=0.9$) relative to the
case of a cloud of radius $L$, when evaluated at phase $\psi\sim \pi/2$.
Instead, 
at small impact parameters ($b\sim 0.1$) 
the Coulomb logarithm becomes comparable 
to that of a cloud of uniform density $\rho^*_0$ and size $L$ (Figure 1,
solid line).


\vskip 25pt
\centerline {6.3 {\it Tidal Distortion}}

\vskip 20pt

Since the vector $(GM/V^2)\vecd$ describes the distortion 
in the density field induced by the satellite during its motion
we can draw   
 the isodensity contours of the perturbed stellar system.
The first series of plots 
describe the deformation of  a density contour  
 lying inside $R$; the second reproduces  the deformation induced by the
passage of the perturber moving within the isodensity contour
in the case of a uniform finite size cloud.

Figure 3 illustrates the evolution of a isodensity level 
in the orbital plane, for  $b/L=1.5.$
The  satellite is moving 
along the trajectory given by equation (44)
 and the displacement vector is computed 
for $GM/bV^2=0.01.$ 
The satellite along its trail  rises a tide and   a bulge
forms which lag behind.

In Figure 4, we draw a sequence of 
contour levels describing the extent of the deformation
when the satellite hits the contour itself: here, $b/L_c=0.5$ and 
$GM/bV^2=0.01$. 
The sharp decrement  in the density behind $M$ 
 represents the region which is adjacent to the overdensity
responsible to the frictional drag.

\vskip 25 pt
\centerline {6.4 {\it Tidal force in shortlived distant encounters}}
\vskip 20pt

At high relative speeds ($V\gg <v^2>^{1/2}$) and large impact parameters
($b\gg <r^2>^{1/2}$, 
the distortion induced on the stellar system acts back to
cause tidal drag. 
By expanding equation [43] in multipoles, we find, to leading order,  

$$\vF_{\Delta}=\vF_{T}=[GM]^2Nm{<r^2>\over V^2}
\left (5(\vecR\cdot \vecn)\,{\vecR\over R^7}-2{\vecn\over R^5}
\right)\eqno (52)$$
resulting in a total energy loss
$\Delta E$ which  coincides with that of Spitzer, once equation (44)
is adopted
for the motion of the satellite.
The force (52) represents the effect of the global tides
on the satellite and has a component along $\vecV$
and a component along $\vecR$, due to the nonuniformities
excited by the tidal interaction.
The radial component is not conservative, i.e., 
it can not be written in terms of 
an effective  potential but 
affects the energy balance equation.
 
In a flyby, the extent of the distortion increases with time. 
The bulge becomes pronounced 
after the satellite has reached  the distance of closet approach and attains
its maximum at a later time $t\sim 1/\omega$;  
the stars will then adjust into a new equilibrium.
The results thus  seem to indicate that   there is a {\it time delay} 
in between the moment of closet approach and 
the moment of maximum deformation of the bulge.
The magnitude of the distortion $D$ can be computed using equation (39)
and is approximately 
$
{(GM/ V^2)}{(D/2L)}\sim
{(V/ <v^2>^{1/2})}{(\epsilon/b)}(b/L-1)^{-1/2}
.$
where we indicate  with $2 L$ the dimension of the galaxy. 

\vfill 
\eject

\centerline  {6.5 {\it Energy dissipation }}
\vskip 20pt

To gain insight into the new result we need to explore further 
the role played  by the 
tidal force. 
Can the global tidal distortion depicted in Figures (4) and (5) 
induce a drag on $M$ , i.e., an effective 
energy and angular momentum  loss 
for the satellite ? 
How does this loss compare  in magnitude to the energy loss by friction ?

We thus  need to estimate the total energy
change (resulting from the friction and tidal components)  along the 
satellite motion. The calculation, outlined in Appendix C, shows that
$\Delta E<0$ for any arbitrary 
stellar distribution,  implying that orbital energy is transferred 
to the stars,  resulting in an effective 
drag for the satellite.  A {\it conservative} term  from the tidal
field is found which does not contribute to  $\Delta E.$
The total energy loss for $M$ decomposes into two terms
$$\Delta E=\Delta E_{+}+\Delta E_{*}.\eqno (53)$$
The first reads 
$$ \Delta E_{+}=-{[GM]^2\over V}Nm\int_{-\infty}^{+\infty}dt {1\over R^2(t)}
\int_{\cal {D}}\,d_3\vr \,\, n_0(r){\vecR(t)\cdot\vr\over \vert 
\vecR(t)-\vr\vert^3},
\eqno (54)$$
where the domain ${\cal {D}}$ is limited to those 
stellar encounters having  $\vert \vecR(t)-\vr\vert>\epsilon$
(we here restore the notation indicating the dependence of $\vecR$ on 
time $t$ for seek of clearness).
The second is 
$$\Delta E_{*}= 
-[GM]^2 Nm {b^2\over 2}\left [\int_{-\infty}^{+\infty}dt {\Omega_t\over
R^3(t)}\right ] 
\,\,\,
\left[ \int_{-\infty}^{+\infty} dt {\lb_t\over R^3(t)}\right ]
\eqno (55)$$
where  $\Omega_t$ is a function of the actual position of the satellite 
$$\Omega_t(R)\equiv\int_{r'<R(t)}d_3\vr'n_0(r');\eqno (56)$$
$\Omega_t$ is proportional to
the stellar mass comprised within a sphere of radius $R(t),$ 
and $\lb=1-\Omega$. 
$\Delta E_{*}$
 gives a nonvanishing contribution
to the energy loss only if the satellite in its past
traveled inside the stellar background.

Restricting the 
analysis to the case of a finite size cloud of radius $L$,
we are able to show  that 
for $b/L>1$ tidal drag leads to an energy loss given by equation (54) 
$$\Delta E=-8\pi {[GM]^2\over V^2}\rho_0
\left [
L-{L^3\over 3b^2}-(b^2-L^2)^{1/2}\sin^{-1}{L\over b}\right ]
\eqno (57)$$

In a close encounter
(i.e., with $b<L$) the frictional force becomes important and the
total energy loss can be written as sum of two contributions: 

$$
\eqalignno {
\Delta E_+= & 
-4\pi{[GM]^2\over 3 V^2}\rho_0
\cr & 
\left [
-{2L^3\over b^2}+{L^2L_M\over b^2} +6L -7L_M-3L_M\ln{2L+L_M\over L_M}
+3L_M\,\ln{L_M/\epsilon}\right ]
& (58)  \cr }$$
and
$$\Delta E_*=-{4\pi\over 3}{[GM]^2\over V^2}\rho_0 
{L_M^3\over 4 b^2}\left [1-{L_M^3\over 8L^3}\right ]
\eqno (59)$$

The last term in the square bracket of equation (58)
 represents the energy 
loss by dynamical friction (from eq. [47] and [49]):
here we show that the relevant scale in the Coulomb logarithm is 
$$b_{max}=L_M=2(L^2-b^2)^{1/2},
\eqno (60)$$
the effective length  traveled by satellite  during its high speed
encounter, within the galaxy. This is the first analytic 
self-consistent derivation
of the maximum impact parameter $b_{max}.$ 
The derivation of equations (57-59)
is outlined  in Appendix C.

Figure 5 depicts the relative contributions to the total energy loss.
When $b>L$ the only energy loss is induces by the
tidal response and is given by equation (57) (solid line).
For $b<L,$ the dot-dashed line refers to the contribution
resulting solely from ``friction" (from eq.[42]).
The dashed line denotes the contribution from equation (59),
while solid line gives the total energy loss.
Clearly, in equation (58) the additional terms resulting from the tidal
response tend to reduce the contribution resulting from friction.
This  goes in the direction indicated by  numerical findings
of Hernquist \& Weinberg (1989; despite  
the misuse of the  definition of the self-gravity
of the response) and Zaritsky \& White (1988).
The decrement increases with decreasing $b/L,$ 
as illustrated in Figure 5.

\vskip 25pt
\centerline {CONCLUSIONS}
\vskip 20pt
We have derived a formalism, within  the linear response theory, 
for the analysis of  the interaction of a massive particle (the perturber)
with a spherical galaxy whose equilibrium is described by
a one-particle distribution function.

We applied the perturbative technique to the case
of a satellite moving at high speed across a stellar system.
We found a natural decomposition of the back-reaction force
into two components: a global component resulting 
from the ``tidal" interaction  and a 
component that is related to  dynamical friction. 

The results are relevant  only to the description of 
shortlived encounters since, in the
application, we neglected the self-gravity of the stellar background.
In paper II (Colpi 1997) 
we will explore the accretion of a satellite 
orbiting around a spherical companion galaxy. 
In a consistent manner we will account for the
self-gravity of stars and for the actual dynamics of the satellite.
We will present a semianalytical 
method for 
the description of the process of tidal capture of a satellite, and
for the analysis of the evolution of a `` binary" with the aim at exploring 
the nature of the gravitational 
interaction between the two systems.
Irreversible processes can cause orbital decay by exciting stars  
in ``near" resonance with the orbit of the satellite; this  phase 
 precedes the final plunge in.
The drag experienced by a satellite moving along an arbitrary orbit inside
the galaxy will be the next problem that we wish to consider.

During final preparation of this work, Nelson \& Tremaine (1997)
submitted a paper where a formal study of the linear response theory
is presented. The authors derive properties of the  response operator
in inhomogeneous dynamical systems and express the
coefficient of dynamical friction in terms of the equilibrium fluctuations,
analogously to our equation (16).
Their theory is developed  in an erbitrary inertial frame.
We have instead chosen to describe tha process of irreversible energy 
transfer between the two components in a frame comoving with the center of
mass of the galaxy (eq.[32]).
As previously discussed, we expect that our formalism is preferable
when an effective one--particle dynamics is adopted for the description
of the unperturbed stellar motion because it automatically takes into account
for the conservation of linear momentum in the scattering process of the 
galaxy whit the satellite.

\vskip 20pt
We  gratefully thank B. Bertotti, F. Combes, G. Gyuk,  and 
I. Wasserman
for useful discussions. This research  was carried out with financial support
from the Italian Ministero dell'Universit\`a e della Ricerca 
Scientifica e Tecnologica.

\def \Liu  {{\cal {L}}}
\def \vF   {{\bf {F}}}
\def \vA1  {\bf {A}_1}
\def \vX  {{\bf {X}}}
\def \vB  {{\bf {B}}}
\def \va  {{\bf {a}}}
\def \vw  {{\bf {w}}}

\def \vri {{\bf {r}}_i}
\def \vr  {{\bf {r}}}
\def \vpi {{\bf {p}}_i}
\def \vp  {{\bf {p}}}

\def \vv  {{\bf {v}}}

\def \vecX {{\bf {X}}}
\def \vecn {{\bf {n}}}

\def \vecs {{\bf {S}}}
\def \vecd {{\bf {D}}}
\def \vcsi {{\bf {\xi}}}
\def \grad {{\bf {\nabla}}}
\def \vecR  {{\bf {R}}}
\def \vecV  {{\bf {V}}}
\def \vb {{\bf {b}}}
\def \lb {{{\bar {\Omega}}}}
\def \vecn {{\bf {n}}}

\vskip 30pt
\centerline {APPENDIX A}
\vskip 25pt
\centerline {{\it 1. Total energy loss}}
\vskip 20pt
In this Appendix 
we sketch the derivation of equation (23) for the energy loss
experienced by the satellite during the encounter, using equation (19):
$$
\Delta E=\int_{-\infty}^{+\infty}
\,dt \,\int_{-\infty}^t\,ds \int d_3\vr\,d_3\vv\,\, \fop  
v^b_s\,T^{ba}\,V_t^a \eqno (A1)
$$
where index $t$ (or $s$) denote that the variable is evaluated at time
$t$ (or $s$).
The domain of integration over time $s$ and $t$ can be
interchanged to give, in an equivalent form 
$$
\Delta E=
\int_{-\infty}^{+\infty}
\,ds \,\int_s^{+\infty}\, dt \int d_3\vr\,d_3\vv \,\,\fop 
 v^b_s\,T^{ba}\,
\left (V^a +v^a- v^a\right ).
\eqno (A2)
$$
Restoring the complete expression for the correlation tensor 
we have
$$\eqalignno {
\Delta E=
&
-[GM]^2 Nm^2\beta \int\,d_3\vr\,\,d_3\vv
\,\,\fop
\cr &
\int_{-\infty}^{+\infty}
\,ds\,
\int_s^{+\infty}\,dt  v_s^b
{R^b_s-r^b_s\over 
\vert \vecR_s-\vr_s\vert^3}
\left[  -{d\over dt}{1\over  
\vert \vecR_t-\vr_t\vert}
+v^a_t\,\,{R_t^a-r_t^a\over 
\vert \vecR_t-\vr_t\vert^3}\right ]:
& (A3)\cr}$$
In the second term of equation (A3) the vectors
$\vecR,\vr$ and $\vv$ 
are evaluated at time $t$.

Because of the isotropy of the unperturbed stellar system,
the contribution to $\Delta E$ resulting from the 
first term in squared bracket,
$$
-[GM]^2 Nm^2\beta \int\,d_3 \vr\,\,d_3\vv
\fop 
\int_{-\infty}^{+\infty}
\,ds\,\, v_s^b
{R^b_s-r^b_s\over 
\vert \vecR_s-\vr_s\vert^4}
\eqno (A4)$$
vanishes identically. The second term gives instead 
equation (23) together with (24).
\vskip 20pt
\centerline {\it 2. Impulse approximation}
\vskip 20pt

In the impulse approximation, the motion of the satellite
and of the background stars
is uniform. We can evaluate $\Delta E$ according
to the simplifying assumption that $\vecR_t=\vb+t\vecV$
and $\vr_t=\vr+t\vv,$ with $\vecV$ and $\vv$ constant 
vectors and $\vb\cdot\vecV=0.$
Defining $\va=\vb-\vr$ and $\vw=\vecV-\vv$, 
the scalar ${\cal {J}}$ is found to be 
$${\cal {J}}=\int_{-\infty}^{+\infty} \,dt\,\,  \vv
\cdot {
\va+t\vw\over \vert \va+t\vw\vert^3}=2\,\vv\cdot \vB
\eqno (A5)$$
with 
$$\vB \equiv {\va_{\perp}\over a^2_{\perp}w}.\eqno (A6)$$
In equation (A6) $\va_\perp$ is the component of $\va$ 
in the orthogonal direction of vector $\vw$ and is equal to
$\va_{\perp}=\va-(\va\cdot\vw)\vw/w^2.$
The total energy loss thus reads
$$\Delta E=-2[GM]^2 Nm^2\beta \int d_3\vr\,\,d_3\vv \fop 
(\vv \cdot \vB)^2.
\eqno (A7)$$

The energy loss by recoil is recovered from equation (A7)
in the limit $b\to \infty$ and $V\to \infty$ valid if
the satellite
moving at high speed maintains always at a distance from the stellar
distribution.
To first leading order, 
the vector $\vB$ reduces to $\vB=\vb/(V\,b^2)$ 
yielding
$$\Delta E=-2[GM]^2 Nm/(V^2\,b^2).\eqno (A8)$$
(We recall that $\int\,d_3\vv \fop (v^a)^2=\sigma^2 n_0(r).$)

Instead, in the limit $V\to \infty$ but $r/b$ finite, 
$$\vB={ \va+(\vr\cdot\vecV)\vecV/V^2
\over
V\left [\va+(\vr\cdot\vecV)\vecV/V^2\right ]},\eqno (A9)$$
and
the energy loss
equals 
$$
\Delta E=-2 {[GM]^2\over V^2}Nm^2\beta \sigma^2
\int d_3\vr\,n_0(r)
{1\over [a^2-(\vr\cdot \vecV)^2/V^2]}.
\eqno (A10)$$
Expanding equation (A10)  to order $<r^2>/b^4$ we 
recover Spitzer result (eq. [28]).

\vskip 30pt

\centerline  {APPENDIX B}
\vskip 25pt

\centerline {\it Wake and global tides}
\vskip 20pt
In this Appendix we summarize the main steps that lead to
the decomposition of the force $\vF_{\Delta}$  in 
terms of $\vF_{DF}$ 
and $\vF_{\Delta}.$  Both components are derived to order
$O(1/V^2)$.  

The density response $\Delta n$ is calculated  from equation (36),
in the high speed limit, according to equations (34) and (35).
The force term (the one in squares brackets)
and  the density field 
$n_0(\vert \vr-(t-s)\vv\vert)$ 
are expanded, in equation (36), to first order in the velocity $\vv,$
contributing  to order $GM/V^2$.

Let us denote with
$\Delta_1 n$ the distortion resulting from the
expansion of the force term, and with $\Delta_2 n$
the one from the expansion of the unperturbed density
$n_0$. The global distortion will
therefore be
$$
\Delta n=\Delta_1 n +\Delta_2 n
\eqno (B1)$$
For ease of analysis, we introduce
the vector $$\vX\equiv
\vecR-\vr -(t-s)\vecV
\eqno (B2)$$
so that
$$
{X^a+(t-s)v^a\over \vert \vX+(t-s)\vv\vert^3}={X^a\over X^3}
+(t-s)v^b{\partial\over \partial X^b}{X^a\over \vert \vX\vert ^3}
+O(v^2) \eqno (B3)$$
to first order in $\vv$ ($\vecX$ has no relation with the one introduced
in $\S 4.1$).

According to equation (36)
$$\eqalignno{
\Delta_1 n=
& 
GMm\beta\left ( {m\beta\over 2\pi}\right )^{3/2}
\,\int_{-\infty}^t\,ds\,\,\int d_3\vv\,\, n_0(r)e^{-{1\over 2}\beta
mv^2}
\cr &
\left \{  
v^a\left [{X^a\over X^3}
-{R^a-(t-s)V^a\over \vert \vecR-(t-s)\vecV\vert^3 }\int_
{r'<d}\,d_3\vr'\,\,n_0(r')\right ]
+(t-s)\, v^a\,v^b {\partial\over \partial X^b}{{X^a\over X^3}}\right \}
& (B4)\cr }$$
where $d$ simply denotes $\vert
\vecR-(t-s)\vecV\vert.$
The term in squared braked vanishes identically because of isotropy.
Only the term $\propto v^a\, v^b$ gives a contribution to the
distortion.
Since for a Maxwellian distribution function
$$\int d_3\vv \,\,e^{-{1\over 2}\beta mv^2}\,\,v^a\,v^b=
{\pi^{3/2}\over 2} \left ({1\over 2}m\beta\right )^{-5/2}\,\delta^{ab}
\eqno (B5)$$
we finally have
$$\Delta_1 n(\vr)=
GM
n_0(r) \,
\,\int_{-\infty}^t\,ds\,\,(t-s) \nabla_{X}\cdot
\left [ {\vX\over \vert \vX\vert^3}\right ].
\eqno (B6)$$
Using the divergence theorem 
we have 
$$\Delta_1 n(\vr)=
4\pi \,GM
n_0(r) \,
\,\int_{-\infty}^t\,ds\,\,(t-s)\delta_3(\vX)
\eqno (B7)$$
which is equivalent to equation (37).

The global distortion in the density field is derived expanding $n_0$
to first order in $\vv;$ using Gauss theorem we derive the following 
expression 
 $$
\eqalignno{
\Delta_2 n(\vr)= &
GMm\beta\left ( {m\beta\over 2\pi}\right )^{3/2}
\left (-{\partial n_0\over \partial r^b}\right )
\,\int_{-\infty}^t\,ds\,\,(t-s)\int d_3\vv e^{-{1\over 2}\beta mv^2}
\cr &
v^a\,v^b\left [{X^a\over X^3}
-{R^a-(t-s)V^a\over \vert \vecR-(t-s)\vecV\vert^3 }\int_
{r'<d}\,d_3\vr'\,\,n_0(r')\right ]
&  (B8)
\cr }$$
Using equation (B5) and introducing the velocity
vector $\vecn=\vecV/V,$  we find that 
 the density response reads
$$
\eqalignno {
\Delta_2 n(\vr) =
& \,-{GM\over V^2}{\partial n\over \partial r^a}
\cr &
\int_0^{+\infty}dx\,\,x\,\,\left [
   {R^a -\vr  -n^ax\over \vert \vecR- \vr -\vecn x\vert^3}
 -{R^a  -n^ax\over \vert \vecR-\vecn x\vert^3}
 +{R^a  -n^ax\over \vert \vecR-\vr-\vecn x \vert^3}
\lb(\vert \vecR-\vecn x\vert)\right ]
& (B9) \cr}$$
where $\lb$ is defined by equation (41).
Equation provides the expression of the displacement vector
in integral form and after lengthy calculations
one  recovers equation (39) for $\vecd$.

\def \vcsi {{\bf {\xi}}}
\def \grad {{\bf {\nabla}}}
\def \vecR  {{\bf {R}}}
\def \vecV  {{\bf {V}}}
\def \vb {{\bf {b}}}
\def \vecn {{\bf {n}}}

\def \de  {{\vert \vecR-\vr\vert^3}}

\vskip 30pt
\centerline {APPENDIX C}
\vskip 25pt
\centerline {1.\it {The total energy loss}}
\vskip 20pt

In this Appendix we estimate the extent of the 
 energy loss 
resulting from friction and from the tides excited by the
satellite during its passage: 
$$\Delta E=\int_{-\infty}^{+\infty}\,dt\,\,\vecV\cdot\vF_{\Delta} 
\eqno (C1)$$
Within  TLR, the macroscopic force on
$M$ is found to be 

$$
\vF_{\Delta}=-GMNm\,\int\,d_3\vr [\Delta_1 n(\vr)+
\Delta_2 n(\vr)] \,
{\vecR- \vr\over \vert \vecR-\vr\vert^3}
\eqno (C2)$$
with $\Delta_1 n$ and $\Delta_2 n$ given by equations (37) and (38).
The contribution coming from $\Delta_1n$ directly gives: 
$$
\vF_{\Delta_1}=-{4\pi (GM)^2Nm\over V^2}\int_0^\infty\,d\tau {n_0(
\vert\vecR -\vecV\tau\vert)\over \tau}
\eqno (C3)
$$
while the term containing $\Delta_2n$ in eq. (C2) can be written as
the sum of three terms:
$$
\eqalignno{
\int&\,d_3\vr \Delta_2 n(\vr)\,{\vecR- \vr\over \vert \vecR -\vr\vert^3}=
-{GM\over V^2}\int d_3\vr
& (C4) \cr
&\left \{ \nabla\cdot\left[\vecd_{\vecR-\vr}n_0(r)
{\vecR- \vr\over \vert \vecR-\vr\vert^3}\right ]
 -n_0(r)
{\vecR- \vr\over \vert \vecR-\vr\vert^3}\nabla\cdot\vecd_{\vecR-\vr}
-n_0(r)\vecd_{\vecR-\vr}\cdot\nabla {\vecR- \vr\over \vert \vecR-\vr\vert^3}
\right\}\cr}
$$
where  $\vecd_{\vecR-\vr}$ is given by equation (39).
The first term vanishes due to the divergence theorem, the second 
exactly cancels the Chandrasekhar term (C3) leaving only the third 
contribution which contains both frictional and tidal effects to give:
$$\vF_{\Delta}^a={[GM]^2\over V^2}Nm\int \,d_3\vr n_0(r) D^b_{(\vecR
-\vr)}\,\,\,
\nabla_{\vecR}{
R^b-r^b\over \de}.\eqno (C5)$$
The force  (C5) can be decomposed into the sum of 
a conservative  contribution plus
a component leading  to dissipation of energy:
$$
\eqalignno{
\vF_{\Delta}=
& 
\nabla_{\vecR}\left [{[GM]^2\over V^2}
Nm\,\int\,d_3\vr\,n_0(r)
\vecd_{(\vecR-\vr)}\,\cdot \,{(\vecR-\vr)\over \de}\right ]
\cr &
-{[GM]^2\over V^2}Nm\int_{\cal {D}} \,d_3\vr\,n_0(r)
{R^b-r^b\over \de}\nabla_{\vecR}\,D^b_{(\vecR-\vr)}.
&  (C6) \cr}$$
For ease of analysis we introduce the vector
$$\vcsi=\vecR-\vr\eqno (C7)$$
expressing the relative distance between the satellite and the stars of 
the background.
In the dissipative component the spatial integration is then limited
to a domain ${\cal {D}}$ satisfying the condition 
$\xi> \epsilon$ (see eq. [27]).  

The vector $\vecd$ 

$$D^b_{\xi}=
\left [ n^b\ln(\xi-\vcsi\cdot\vecn)+{
\xi^b-n^b(\vcsi\cdot\vecn)\over
\xi-\vcsi\cdot\vecn}\right ]
-\left [
 n^b\ln(R-\vecR\cdot\vecn)+{
R^b-n^b(\vecR\cdot\vecn)\over
R-\vecR\cdot\vecn}\right ]
 +S^b
\eqno (C8)
$$
can be schematically written in the form
$$
D^b_{\xi}=\left [ \vecn,\vcsi \right ]^b-
\left [ \vecn,\vecR \right ]^b +S^b
\eqno (C9)
$$
where  $\vecs$ is defined by equation (40).

In order to calculate the total energy loss we need to
evaluate explicitly the component of the force along the
velocity unit  vector $\vecn$.
Along the unit vector $\vecn,$ the force is given 
$$
n^a\,F^a_{\Delta}= -{[GM]^2\over V^2}
Nm\int \, d_3\vr\,\,n_0(r) n^a{\xi^b\over \xi^3}
{\partial D^b\over \partial R^a}
\eqno (C10)
$$
One can prove that
$$n^a{\xi^b\over \xi^3}{\partial 
\over \partial \xi^a}\left [ \vecn,\vcsi \right ]^b={
1\over \xi^3}
\eqno (C11)$$
and 
$$n^a{\xi^b
\over \xi^3}{\partial \over \partial R^a}\left [ \vecn,\vecR \right ]^b={
\xi^b\over \xi^3}{R^b-n^b \,R\over
R(R-\vecn\cdot\vecR)}
\eqno (C12) $$

Substituting (C11) and (C12) in (C10), and 
using  Gauss theorem, we find 

$$n^a\,F^a_{\Delta}=-{[GM]^2\over V^2} Nm\left \{
-{\Omega(R)\over R^3}
+n^a{\partial S^b\over \partial R^a}{R^b\over R^3}\Omega(R)\right \}+
\int d_3\vr\,\,{n_0(r)\over \de}.
\eqno (C13)$$
Noting that $${\Omega (R)\over R^3}\equiv {1\over R^3}\int_{r<R}
d_3\vr \,\,n_0(r)=\int \,d_3\vr {n_0(r)\over \de}
-{1\over R^2}\int d_3\vr\,\,n_0(r){
\vecR\cdot\vr\over \de}
\eqno (C14)$$
we find that $\Delta E$ splits into two terms $\Delta E_+ + \Delta E_*$ 
given by

$$\Delta E_+=-{[GM]^2\over V}Nm
\int_{-\infty}^{+\infty}\,dt 
{1\over R^2}\int\,d_3\vr
\,\,n_0(r){\vecR\cdot\vr\over \de}
\eqno (C15)$$
and $$\Delta E_*=
-{[GM]^2\over V}Nm\int_{-\infty}^{+\infty}
\,dt \,\,
n^a\,\,{\partial S^b\over \partial R^a}{R^b\over R^3}\Omega(R).
\eqno (C16)$$
Both terms are negative as shown below.

As regard to the first term $\Delta E_+$ (C15), 
owing to the isotropy of the background,
we can arbitrarily select a direction for $\vecR$ (along $z$ axis)
so that the resulting volume integral
$$\int d_3\vr\,\, n_0(r){R\,z\over (R^2+r^2)^{3/2}}
[1-
2Rz/(R^2+r^2)]^{-3/2}\eqno (C17)$$
has a kernel of the form $(1-z)^{-3/2}=[1+3z/2-15z^2/8+ \,\,
O(z^3)]$
when expanded in series. Since only the terms with odd powers
give a nonvanishing contribution, 
the volume integral has definite positive sign, resulting in a
net energy loss.
Also the term in equation (C16) can be suitably written in a form
involving only positive functions.

We know evaluate $\Delta E_*$.
If we explicitly insert the expression of 
$\vecs$ as given by equations (40) and (41) into equation (C16)
we have that the integral

$$I\equiv {\Omega(R) \over R^3}n^aR^b{\partial \over \partial R^a}
\int_0^{+\infty} dx \,\,x\,\,{R^b-n^bx\over
\vert \vecR-\vecn\,x\vert^3}\lb(\vert \vecR-\vecn\,x\vert )
\eqno (C18)$$
is equal to
$$I=-{\Omega(R) \over R^3}R^b\int_0^{+\infty}\,dx\,\,x {\partial\over\partial x}
\left [ {R^b-n^bx\over
\vert \vecR-\vecn\,x\vert^3}\lb(\vert \vecR-\vecn\,x
\vert )\right ]\eqno (C19)$$
giving a contribution to the energy loss
$$
\Delta E_*=
-[GM]^2Nm\int_{-\infty}^{+\infty}
\,dt {\Omega (R_t)\over R^3_t}\int_0^{\infty}\,ds {R^2_t-
\vecR_t\cdot \vecV s
\over \vert \vecR_t-\vecV\,s\vert ^3}\lb(\vert
\vecR_t-\vecV\,s \vert )\eqno (C20)$$
where  we recovered the
dependence upon $t$ of 
$\vecR_t,$ denoting the satellite position at current time $t$.

Since, in a high speed encounter, the satellite moves along a trajectory
that can be approximated as linear, we can evaluate explicitly the
integral over the variable $s$ using equation (44).
Noting that the following equality holds
$$\int_{-\infty}^{+\infty}\,dt {\lb(R_t)\over R_t^3}=
\int_{-\infty}^{+\infty}\,dt {[1-\Omega (R_t)]\over R_t^3}={2\over b^2V}-
\int_{-\infty}^{+\infty}\,dt {\Omega (R_t)\over R_t^3}
\eqno (C21)$$
after some  calculation we find that  
$$\Delta E_*={[GM]^2\over V}Nm \left [
{b^2V\over 2}\left ( \int_{-\infty}^{+\infty}\,dt {\lb\over R_t^3}
\right )^2-\int_{-\infty}^{+\infty}\,dt {\lb\over R_t^3}\right ].
\eqno (C22)$$
Using equation (C21) we recover eq.(55).

\vskip 25pt
\centerline {2. {\it Energy loss in a uniform cloud}}
\vskip 20pt
Here we provide a quantitative estimate of the total energy loss
and determine the relative importance of the tidal and frictional 
contributions to the process of dissipation. We consider the 
case of a uniform cloud of mass density $\rho_0=Nm n_0$ and radius $L.$

\noindent
{\it CASE A}: If the satellite travels outside the spherical cloud
the only contribution to the energy loss comes from the tidal field
and is given by 
$$\Delta E_+=
-{[GM]^2\over V}\rho_0\int_{-\infty}^{+\infty}
\,dt 
\left [ -{4\pi\over 3} {L^3\over R_t^3} -4\pi{L\over R_t} +
2\pi \ln {R_t+L\over R_t-L}\right ]
\eqno (C23)$$
Since $R_t=(b^2+(Vt)^2)^{1/2}$,
the integral over time $t$ can be evaluated analytically yielding
$$\Delta E_+=-8\pi {[GM]^2\over V^2}\rho_0
\left [
L-{L^3\over 3b^2}-(b^2-L^2)^{1/2}\sin^{-1}{L\over b}\right ]
\eqno (C24)$$
where $b$ is the impact parameter ($b>L$).
In the limit $b\gg L$ equation (C24) reduces to (28).

\medskip
\noindent
{\it CASE B}:
If during the shortlived encounter the satellite enters the spherical cloud
($b<L$), the total energy loss is given by equation (54) plus (55).
Notice that the satellite travels  a distance $L_M=2 (L^2-b^2)^{1/2}$, 
within the cloud, over a time $\tau_{e}=L_M/V$.
In this case
$$\Delta E_{+}=-{[GM]^2\over V^2}\rho_0\left [
\int_{-L_{M}/2}^{L_{M}/2}
\,d(Vt) \left ( I_{<}+I_{>}\right )+2\int_{-\infty}^{-L_{M}/2}
d(Vt) I_o\right ]
\eqno (C25)$$
where $I_{<}$ 
$\,\,$ 
($I_{>}$) is the volume integral resulting from
the region of the cloud with $[0<r<R-\epsilon]$ 
\break
($[L>r>R+\epsilon]$ respectively):
$$I_{<}+I_{>}=-{16
\pi \over 3}+2\pi\ln{L^2-R^2_t\over \epsilon^2}
\eqno (C26)$$
while
$$I_o=-{4\pi\over 3}{L^3\over R^3}
-4\pi {L\over R}+2\pi \ln{R+L\over R-L}\eqno (C27)$$
The time integrals can then be carried out analytically:
$$
\eqalignno{
\Delta E_+= &
-4\pi{[GM]^2\over 3 V^2}\rho_0
\cr &
\left [
-{2L^3\over b^2}+{L^2L_M\over b^2} +6L -7L_M-3L_M\ln{2L+L_M\over L_M}
-3L_M\ln{\epsilon\over L_M}\right ]
&  (C28) \cr}$$

Similarity, we can carry out the integration of equation (55)
involving the functions $\Omega $ and $\lb.$
The calculation is simple and lead to the following expression for
$$\Delta E_*=-{4\pi\over 3}{[GM]^2\over V^2}\rho_0 
{L_M^3\over 4 b^2}\left [1-{L_M^3\over 8L^3}\right ]
\eqno (C29)$$
where $L_M=2\sqrt{L^2-b^2}$ is the length of the portion of trajectory
within the cloud.

\vskip 30pt
\centerline {REFERENCES}
\vskip 20pt

\def \ref {\par\noindent\hangindent=1.5truecm}

\ref Barnes, J.E., \& Hernquist, L. 1992, ARA\&A, 30, 705

\ref Begelman, M.C., Blandford, R.D., \& Rees, M.J. 1980, Nature, 287, 307

\ref Bekenstein, J.D., \& Maoz, E. 1992, ApJ 390, 79

\ref Binney, J., \& Tremaine, S. 1987, Galactic Dynamics, Princeton Univ. Press

\ref Bontekoe, Tj. R., \& van Albada, T.S. 1987, MNRAS, 224,349

\ref Chandrasekhar, S. 1943, ApJ, 97, 251

\ref Colpi, M. 1997, submitted to ApJ
 
\ref Duncan, M.J., Faruki, R.T., \& Shapiro, S.L. 1983, ApJ, 271, 22

\ref Hernquist, L., \& Weinberg, M.D. 1989, MNRAS, 238, 407 

\ref Kandrup, H.E. 1981, ApJ, 244, 316

\ref Kandrup, H.E. 1983, Astroph. Space Sci., 97, 435

\ref Lin, D.N.C., \& Tremaine, S. 1983, ApJ, 264, 364

\ref Lynden-Bell, D., \& Kalnajs, A. 1972, MNRAS, 157, 1

\ref Maoz, E. 1993, MNRAS, 263, 75


\ref Mulder, W.A. 1983, A\&A, 117, 9

\ref Nelson , R.W., \& Tremaine, S. 1997, astro-ph/9707161, 14 july
 
\ref Quinn, P.J., \& Goodman, J. 1986, ApJ, 309, 472

\ref Spitzer, L. 1958, ApJ, 127, 17

\ref Tremaine, S., \& Weinberg, M.D. 1984, MNRAS, 209, 729

\ref Weinberg, M.D. 1986, ApJ, 300, 93

\ref White, S.D.M. 1976, MNRAS, 174, 467

\ref White, S.D.M. 1983, ApJ, 274, 53

\ref Zaritsky, D., \& White, S.D.M. 1988, MNRAS, 235, 289

\vskip 30pt
\centerline {Figure Caption}
\vskip 20pt
\parindent=0pt

{\bf Figure 1}: Effective Coulomb logarithm versus phase $\psi$
for a uniform finite size cloud 
{\it Solid} line is for $b=0.1$; {\it dashed} line for $b=0.5$
and {\it dashed-dotted} line for $b=0.9$.

{\bf Figure 2}: Effective Coulomb logarithm  ({\it solid} line; eq.[51])
against phase $\psi$
for $b=0.9$ (right panel) and for $b=0.5$ (left panel); $\epsilon/L=0.01$.
The spherical stellar background has  density profile given by
eq. [50] with $\gamma=3/2$. {\it Dashed} line gives the comparison with
the case  of a uniform 
density cloud.

{\bf Figure 3}
Isodensity contours, in the orbital plane ($x,y$) for $b/L=1.5$ :
{\it Square} denotes the position of the satellite moving along
the $x$ axis, from left to right. The phase $\psi$ decreases as
time elapses and the four panels refer to
$\psi=120,90,60,45$ degrees respectively.

{\bf Figure 4}
Isodensity contours, in the orbital plane ($x,y$) for $b/L=0.5$ :
{\it Square} denotes the position of the satellite moving along
the $x$ axis, from left to right. The phase $\psi$ decreases as
time elapses and the four panels refer to
$\psi=157,130,50,26$ degrees respectively.


{\bf Figure 5}
Energy loss $-\Delta E$ in units of $[GM]^2\rho_0\,L/V^2$ as a function
of $b/L$. The spherical system is uniform and has size $L$.
For $b/L>1$ only tides cause dissipation of energy (eq. [57]-{\it solid} line)
For $b/L<1$ the tidal contribution of equation (59) is indicated by the 
{\it dashed}
and the {\it dot-dashed} line denotes  the contribution from 
the logarithmic term of equation (58) with $\epsilon/L=0.01$. The {\it solid } line gives
the total energy loss.

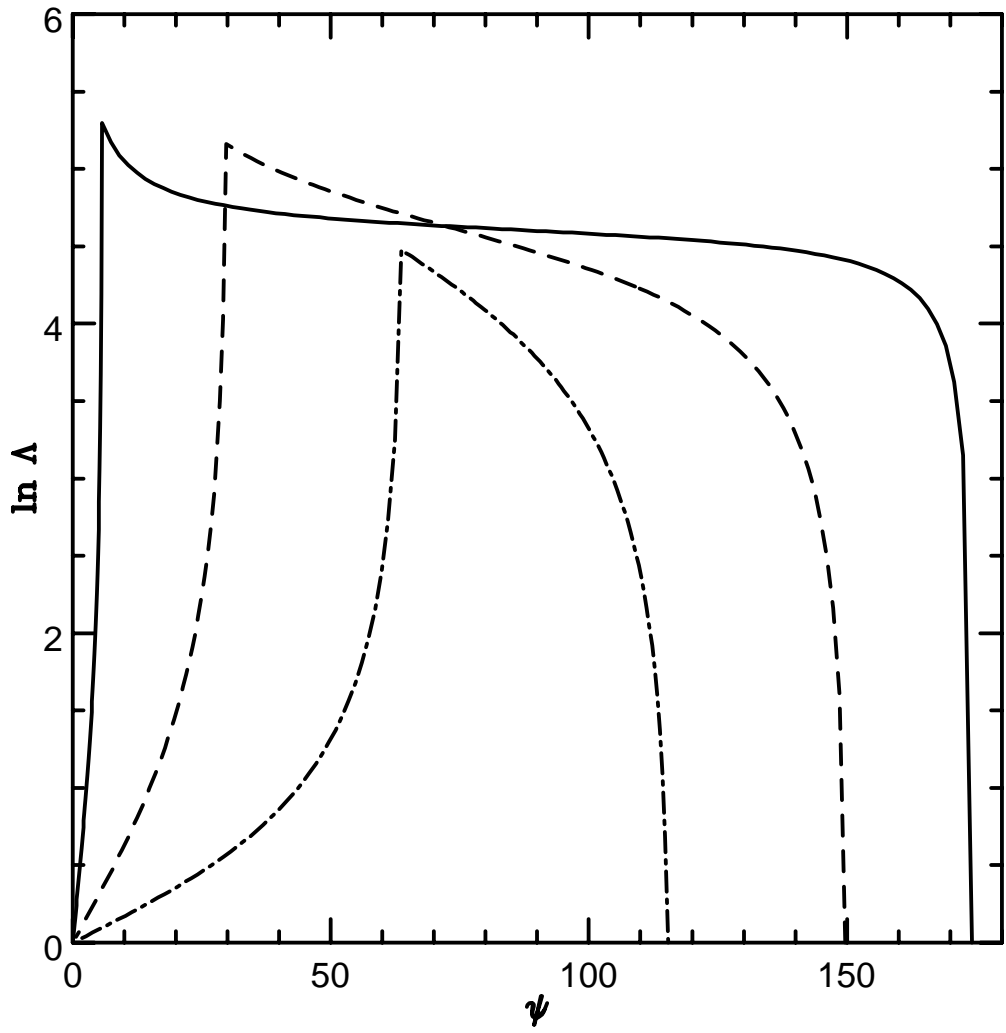
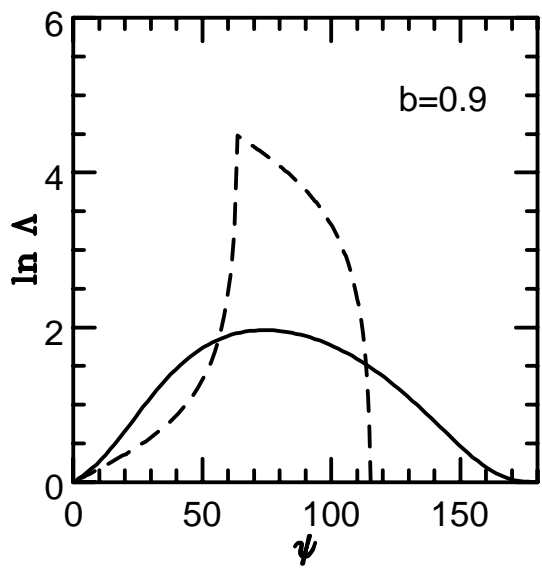
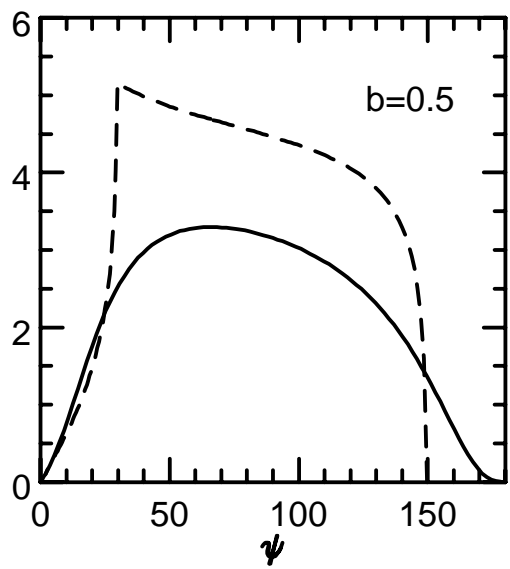
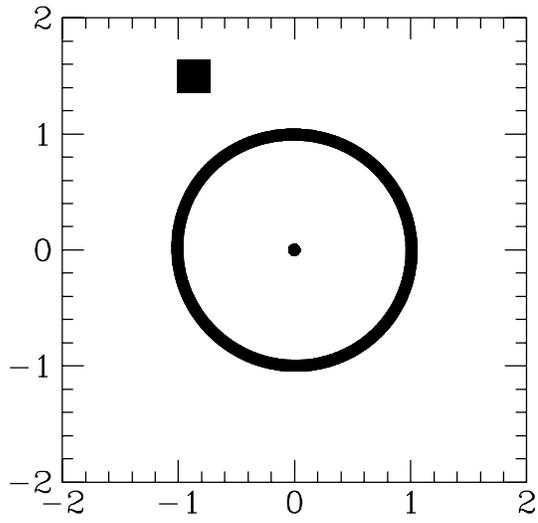
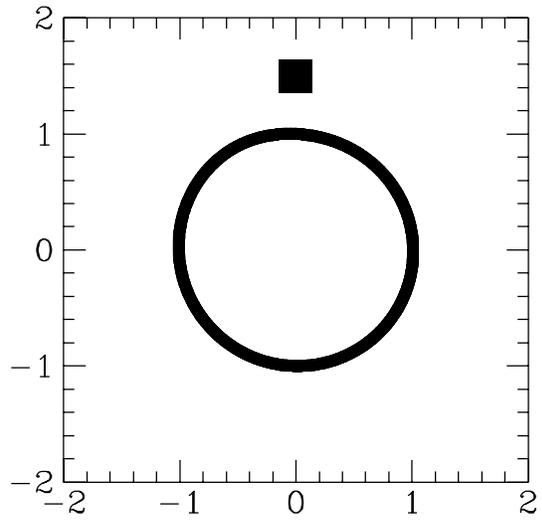
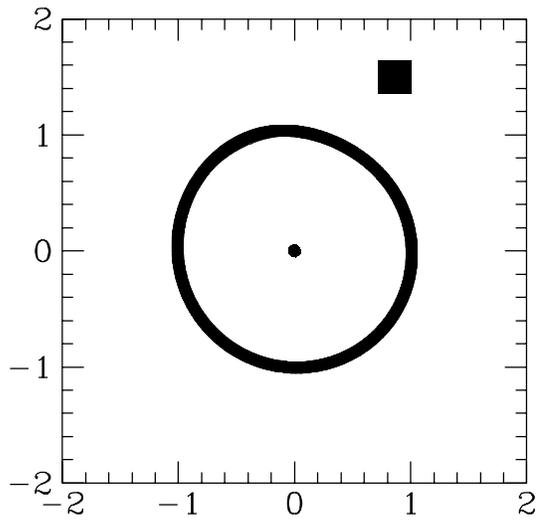
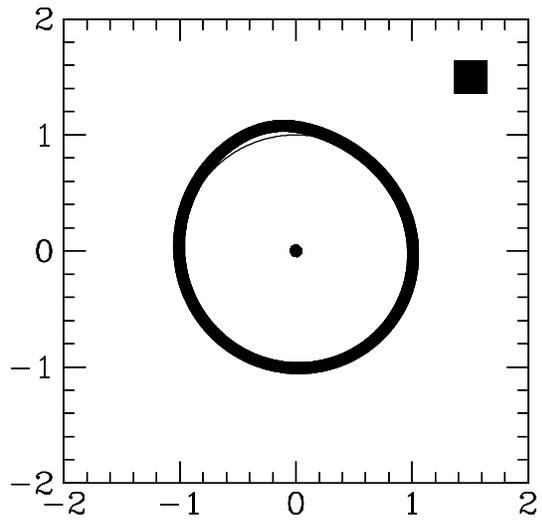
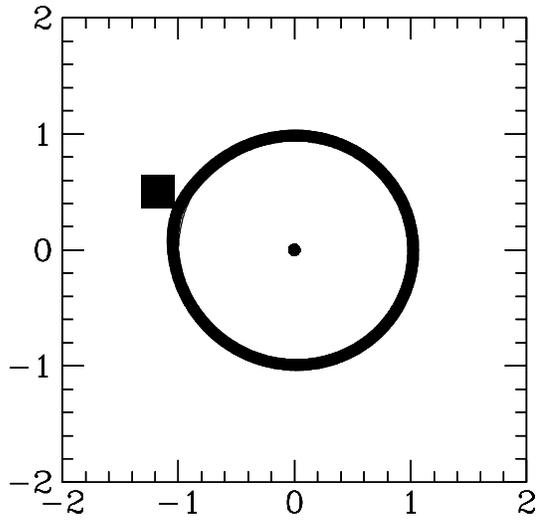
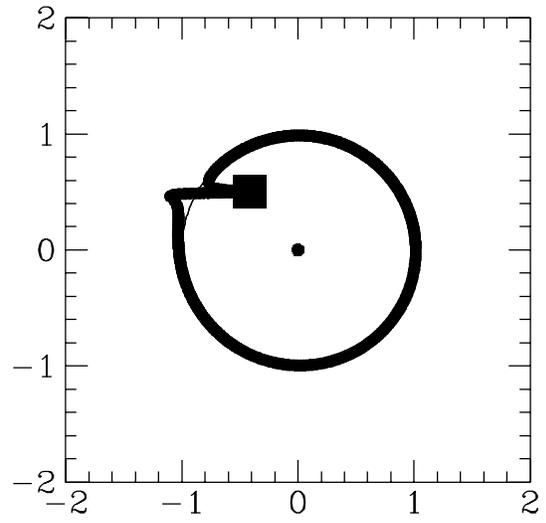
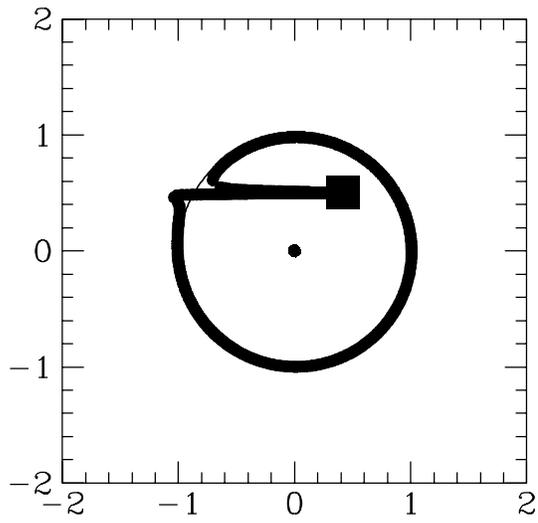
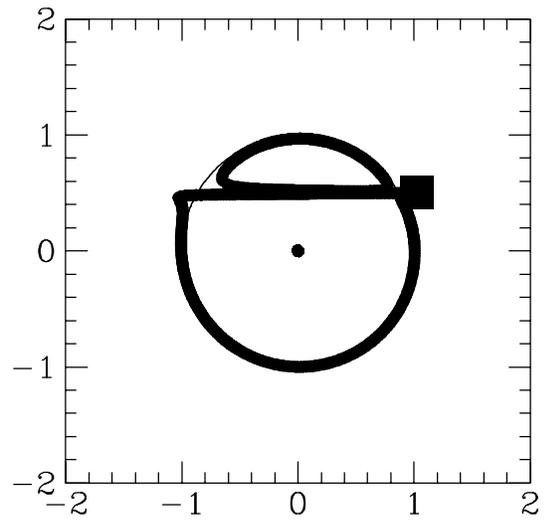
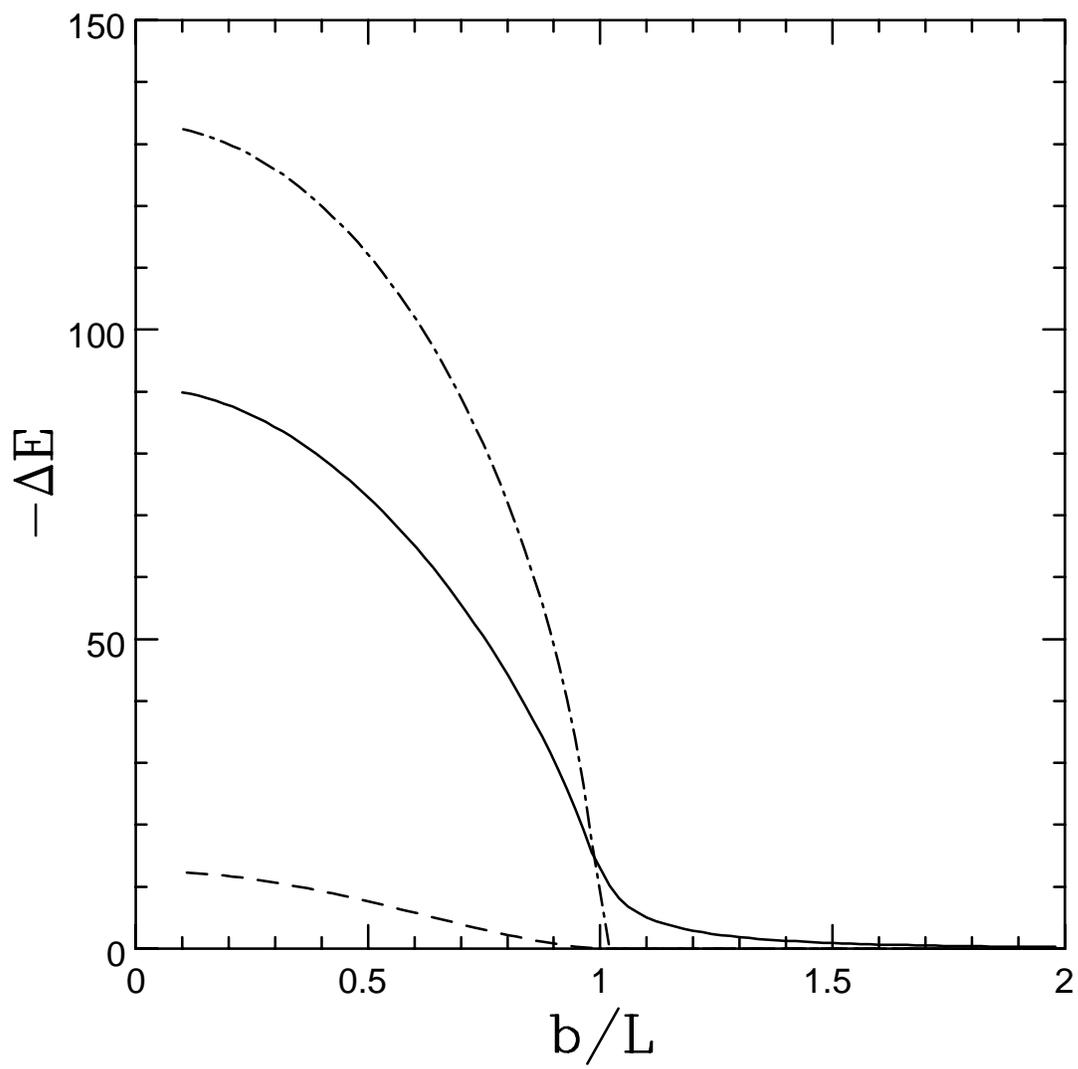
\bye